\documentclass[journal,draftclsnofoot,onecolumn,12pt]{IEEEtran}
\usepackage{amsthm,amssymb,graphicx,multirow,amsmath,color,amsfonts}
\usepackage[update,prepend]{epstopdf}
\usepackage[noadjust]{cite}
\usepackage[latin1]{inputenc}
\usepackage{tikz}
\usepackage{bbm} 
\usepackage{pdfpages}
\usepackage{multirow}
\usepackage{subfig}
\usepackage{comment}
\def\nb0{{\mathbf{0}}}
\def\nb1{{\mathbf{1}}}

\newtheorem{lemma}{Lemma}

\newtheorem{definition}{Definition}

\newtheorem{theorem}{Theorem}
\newtheorem{prop}{Proposition}

\newtheorem{remark}{Remark}

\allowdisplaybreaks 

\usepackage{setspace}	

\begin{document}
\graphicspath{{./Figures/}}
\title{
Exploiting Randomly-located Blockages for Large-Scale Deployment of Intelligent Surfaces
}
\author{
Mustafa A. Kishk, {\em Member, IEEE} and Mohamed-Slim Alouini, {\em Fellow, IEEE}
\thanks{Mustafa A. Kishk and Mohamed-Slim Alouini are with King Abdullah University of Science and Technology (KAUST), CEMSE division, Thuwal 23955-6900, Saudi Arabia (e-mail: mustafa.kishk@kaust.edu.sa; slim.alouini@kaust.edu.sa). This work was funded in part by the Center of Excellence for NEOM Research at KAUST.
} 
\vspace{-4mm}}

\maketitle

\begin{abstract}
One of the promising technologies for the next generation wireless networks is the reconfigurable intelligent surfaces (RISs). This technology provides planar surfaces the capability to manipulate the reflected waves of impinging signals, which leads to a more controllable wireless environment. One potential use case of such technology is providing indirect line-of-sight (LoS) links between mobile users and base stations (BSs) which do not have direct LoS channels. Objects that act as blockages for the communication links, such as buildings or trees, can be equipped with RISs to enhance the coverage probability of the cellular network through providing extra indirect LoS-links. In this paper, we use tools from stochastic geometry to study the effect of large-scale deployment of RISs on the performance of cellular networks. In particular, we model the blockages using the line Boolean model. For this setup, we study how equipping a subset of the blockages with RISs will enhance the performance of the cellular network. We first derive the ratio of the {\em blind-spots} to the total area. Next, we derive the probability that a typical mobile user associates with a BS using an RIS. Finally, we derive the probability distribution of the path-loss between the typical user and its associated BS. We draw multiple useful system-level insights from the proposed analysis. For instance, we show that deployment of RISs highly improves the coverage regions of the BSs. Furthermore, we show that to ensure that the ratio of blind-spots to the total area is below $10^{-5}$, the required density of RISs increases from just $6$ RISs/km$^2$ when the density of the blockages is $300$ blockage/km$^2$ to $490$ RISs/km$^2$ when the density of the blockages is $700$ blockage/km$^2$. 
\end{abstract}
\section{Introduction}
One of the main challenges in the next generation wireless networks is keeping up with the increasing demand for higher data rates. Another concern, arising from the significant increase in the number of cellular-connected devices, is the energy efficiency. Among many potential technologies that can be used to tackle these two challenges, reconfigurable intelligent surfaces (RISs) has stood out. RIS is a surface composed of a number of reflective elements with adjustable phase shifts, lending the capability to modify the impinging signals and steer the reflected waves towards any intended direction~\cite{chen2019channel,mu2019exploiting}. The RIS operation relies mainly on controlling the phase shifts of the reflecting elements, which can be tuned through a simple controller. Hence, compared to a relay that can provide similar job, an RIS consumes much less energy. In addition, according to recent studies in~\cite{ntontin2019reconfigurable,Bjornson_2019}, RISs can outperform relays in some specific scenarios.

The capabilities of RISs can be useful in multiple scenarios, such as signal strength enhancement, interference engineering, and ensuring signal secrecy at eavesdroppers~\cite{basar2019wireless}. One of the scenarios that will highly benefit from the deployment of RISs is providing indirect line-of-sight (LoS) paths for blocked links. In particular, for a user-BS link that is obstructed with a blockage, if there exists an RIS that has an LoS with each of them, the RIS can provide an {\em indirect LoS link} between the user and the BS. Hence, large-scale deployment of RISs in cellular networks has the potential to significantly increase the coverage area of the BSs and reduce the {\em blind-spot} areas, where there are no LoS with any BS. This is particularly important at high frequency bands where signal attenuates severely due to blockages.

Based on the above discussion, the idea of large-scale deployment of RISs has been proposed in the literature. This can be achieved by coating already existing objects, such as buildings or trees, with RISs. In other words, objects that normally act as blockages for communication links can be exploited to enhance the performance of the cellular network. In this paper, we use tools from stochastic geometry to study the potential gain from coating blockages with RISs in terms of enhancing the LoS probability and reducing the average path-loss. To the best of our knowledge, this is the first work to provide an analytical framework for the large-scale deployment of RISs in cellular networks with emphasis on the LoS probability and average path-loss improvement. More details about the contributions of this paper will be provided in Sec.~\ref{sec:cont}. First, we enlist the related work in the next subsection.
\subsection{Related Work}
In this section, we discuss the most related works in literature, which can be categorized into: (i) RIS-enabled communication system analysis and design, (ii) stochastic geometry-based analysis of blockages in cellular networks, and (iii) stochastic geometry-based analysis of RIS-enabled communication systems.

{\em RIS-enabled communication system analysis and design.} Authors in~\cite{8056966} introduced the idea of utilizing metallic reflectors in indoor environments to enhance the communication performance. However, the performance of such setup would be limited by Snell's law where incident and reflection angles should be equal. RISs, on the other hand, have much higher capability to control the reflected waves~\cite{basar2019wireless,zhao2019survey,LIASKOS20191,Liaskos2018,8466374,huang2019holographic,liang2019large,di2019smart,qingqing2019towards}. Authors in~\cite{LIASKOS20191,Liaskos2018,8466374} envision an indoor wireless system where walls are coated with RISs, lending the ability to tailor the wireless environment based on the user's needs, whether it is higher data rate, better secrecy, or wireless charging. However, as discussed in~\cite{huang2019holographic}, RISs have a wide set of applications for enhancing signal quality in the outdoor environment as well. For instance, authors in~\cite{liang2019large,di2019smart} studied the use of RISs for improving downlink transmission in cellular networks. Similarly, authors in~\cite{qingqing2019towards} discussed the importance of RISs for improving coverage at edge-users. As stated already, RISs can be useful in multiple scenarios. For instance, in~\cite{wu2019joint,shi2019enhanced,pan2019intelligent,tang2019joint}, performance analysis of wireless power transmission using RISs is provided. In~\cite{wu2019joint,8723525,8847342,xu2019resource,yu2019enabling}, secrecy analysis of RIS-enabled communication systems is provided. Authors in~\cite{he2019adaptive,he2019large} considered a communication system where RISs are used to enhance localization accuracy. In~\cite{zhang2019reflections}, authors study a system where an RIS-equipped drone is used to enhance wireless coverage.

{\em Stochastic geometry-based analysis of blockages.} The analysis of LoS probability in cellular networks with randomly located blockages was provided in~\cite{6840343}. The authors modeled the blockages using the boolean model where the midpoints of the blockages are modeled as a Poisson point process (PPP) while the length, width, and orientation of the blockages are assumed to be uniformly distributed. These results were later used in~\cite{6857368} to study the Macro diversity in cellular networks where the user associates with the nearest LoS BS within a specific range. However, to maintain tractability, in all these works, the correlation between blockages of different links was ignored. Recent works in~\cite{8114332,8485965,8493525,7982685}, have provided analytical frameworks to capture this correlation. In all the discussed works, the blockages where just considered to study the LoS probability. However, the signals reflected from these blockages towards the receiver were ignored. Recently, authors in~\cite{8254903} studied similar setups while considering the signals reflected from the blockages. Authors in~\cite{o2019mathematical,8839839} also considered similar setup with emphasis on the performance of localization in cellular networks. Given that reflections are limited by Snell's law, for a given user-BS pair, the authors derived the set of locations where placing a reflector with a given length and orientation will reflect the BS's transmitted signals towards the receiver.

{\em Stochastic geometry-based analysis of RIS-enabled communication systems.} As stated, the works that captured the effect of reflected signals focused on using typical metallic reflectors where the reflected signals abide by Snell's law. RISs, on the other hand, have the capability to reflect the signals towards a wider set of directions due to its special characteristics. Multiple works in literature have considered a setup where there exists multiple RISs in the system~\cite{8948323,jung2019optimality,yang2020energy,chaccour2020risk}. However, all these works assume the number of RISs are predefined and their locations are fixed. Stochastic geometry-based modeling of the locations of RISs, on the other hand, is more scarce in literature. In fact, the only existing work, to the best of the authors' knowledge, is~\cite{di2019reflection}. Authors in~\cite{di2019reflection} considered a setup where RISs are modeled using boolean model and derived the probability that a given RIS is capable to provide an indirect path for a given transmitter-receiver pair (the reflection probability). One of the main takes of this work is that this probability is not function of the size of the RIS. This is due to the assumption that the RIS is capable to reflect an impinging wave towards any direction. 

\subsection{Contributions}\label{sec:cont}
In this paper, we use tools from stochastic geometry to analyze a cellular network that suffers from the existence of randomly located/oriented blockages, with subset of these blockages coated with RISs. While the LoS probability has been derived in literature for this setup in~\cite{6840343}, we study the effect of coating a subset of the blockages with RISs on the LoS probability and the average path-loss. More details are provided next.

{\em Large scale deployment of RISs}. This work constitutes the first attempt to study and analyze a cellular network with large-scale deployment of RISs. Aligning with literature, we consider line boolean model for the blockages. Furthermore, we assume that a fraction of these blockages are coated with RISs. We use this framework to study the performance gains when the RISs are used to provide indirect LoS links for user-BS blocked links. However, the proposed framework can be used in future work to study many further applications of RISs such as secrecy enhancement, localization, and interference engineering.

{\em Performance Analysis}. For the explained system setup, we derive multiple useful performance metrics. Namely, we derive the LoS probability between a user and a BS at a given distance. We show how the LoS probability is improved due to coating a fraction of the blockages with RISs. Next, we derive the fraction of the area where users have no LoS with any BSs (blind-spots). We also derive an upper bound for the RIS deployment efficiency, which is the fraction of the RISs that are actually being utilized to provide indirect LoS paths. Finally, we derive the probability that the average path-loss is below a predefined threshold. We show how this probability is improved by  increasing each of the number of the meta-surfaces per RIS and the fraction of RIS-coated blockages.

{\em System-level insights}. With the aid of the derived performance metrics and the numerical results, we draw various system-level insights. For instance, we show that environments with low blockage density ($300$ km$^{-2}$) require small fraction ($2\%$) of RIS-coated blockages to significantly reduce the blind-spot area. However, this fraction considerably increases ($70\%$) at environments with high blockage density ($700$ km$^{-2}$). Furthermore, we use the derived upper bound on the deployment efficiency to emphasize on the importance of well-planned deployment of RISs. In particular, we show that deploying RISs at strategic locations can reduce the required deployment density by more than $70\%$ at high-blockage-density environments and by more than $80\%$ at low-blockage-density environments. Finally, we show that the required fraction of RIS-equipped blockages significantly reduces when we increase the number of meta-surfaces per RIS.
\section{System Model}\label{sec:sys}
We consider a cellular network where the locations of the BSs and the users are modeled as two independent homogeneous PPPs $\Psi_{BS}=\{y_i\}\in\mathbb{R}^2$ with density $\lambda_{BS}$ and $\Psi_{u}=\{u_i\}\in\mathbb{R}^2$ with density $\lambda_{u}$, respectively. The blockages are modeled using line boolean model~\cite{6840343}. In particular, the blockages are modeled as line segments with length $L$ and angle $\theta_b$. The locations of the midpoints of the blockages are modeled as a PPP $\Psi_b=\{z_i\}\in\mathbb{R}^2$ with density $\lambda_b$. For a given blockage whose midpoint located at $z_i$, the value of $L_i$ is uniformly distributed between $L_{\rm min}$ and $L_{\rm max}$. The value of $\theta_{b,i}$ represents the angle between the the blockage and the positive direction of the $x$-axis, and is assumed uniformly distributed between $0$ and $2\pi$. 

A subset $\Psi_R\subset\Psi_b$ of the blockages are equipped with RISs. The density of $\Psi_R$ is $\lambda_R=\mu\lambda_b$, where $0\leq\mu\leq 1$. The value of $\mu$ represents the fraction of the blockages that are equipped with RISs. This model abstracts the proposed direction of implementing RISs on the sides or the fronts of the buildings to enhance coverage. Given that there might be some restrictions on the deployment of RISs on building sides, such as restricting the deployment to the back facades, we assume that the RIS is deployed on only one of the two sides of the line segments that represent the blockages.

Before describing the main performance metrics considered in this paper, we first define some terminologies that will be frequently used throughout the paper.
\begin{definition}[Direct LoS-link]
A direct LoS-link exists when there are no blockages obstructing the path between the user and the BS.
\end{definition}

\begin{definition}[Indirect LoS-link]
An indirect LoS-link exists when there are no blockages obstructing neither the path between the user and the RIS nor the path between the RIS and the BS.
\end{definition}

\begin{figure}
\centering
\includegraphics[width=1\columnwidth]{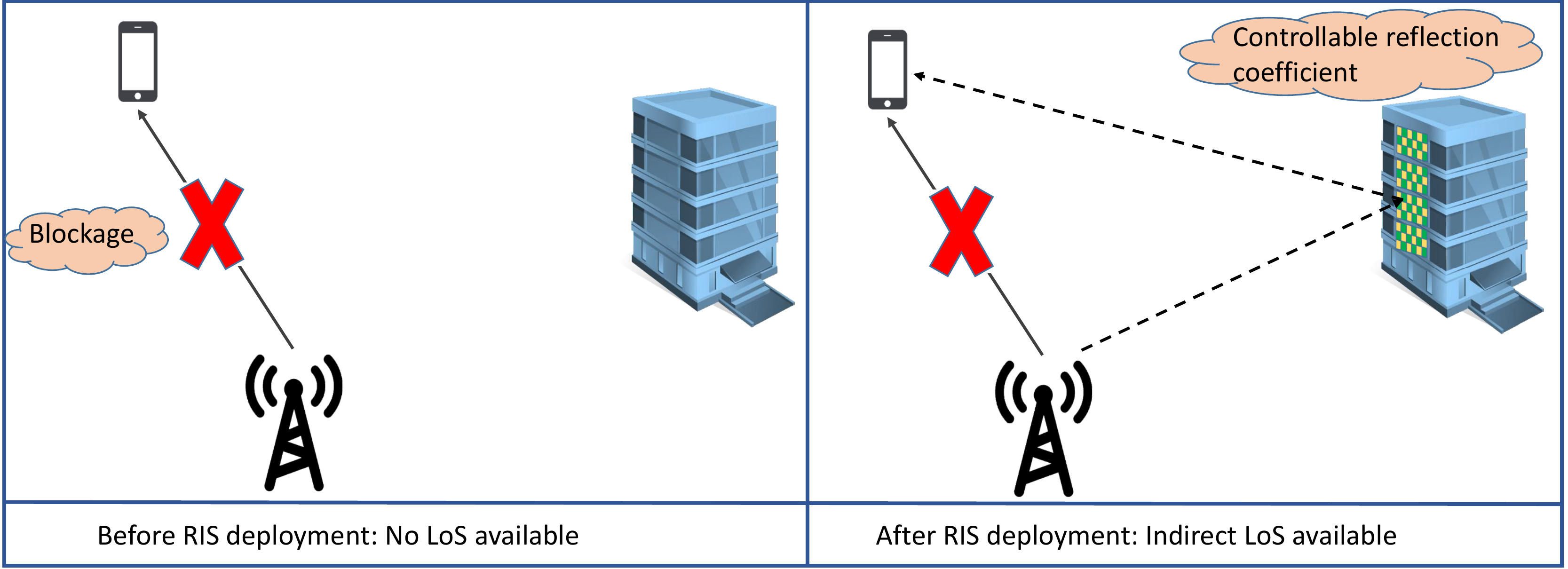}
\caption{ Equipping the building with RIS enables an indirect LoS between the user and the BS.}
\label{fig:motivation}
\end{figure}

\begin{definition}[Blind-Spots]\label{def:not}
blind-spots are the areas that have neither direct nor indirect LoS-links to any BS.
\end{definition}
\begin{table}[t]\caption{Table of notations}
\centering
\begin{center}
\resizebox{\textwidth}{!}{
\renewcommand{\arraystretch}{1.4}
    \begin{tabular}{ {c} | {c} }
    \hline
        \hline
    \textbf{Notation} & \textbf{Description} \\ \hline
    $\Psi_u$; $\Psi_{BS}$; $\Psi_{b}$; & The PPP modeling the locations of the users; the BSs; the midpoints of the blockages \\ \hline
    $\lambda_u$; $\lambda_{BS}$; $\lambda_b$ & The density of $\Psi_u$; $\Psi_{BS}$; $\Psi_{b}$  \\ \hline
    $\Psi_R$& The PPP modeling the locations of the midpoints of the RIS-equipped blockages \\ \hline            
        $\Psi_{R,k}$ & The PPP modeling the locations of the midpoints of the RIS-equipped blockages with $k$ meta-surfaces. \\ \hline  
        $\mathcal{M}$ & The set of possible values for the number of meta-surfaces per RIS $M$. Note that $\Psi_R=\bigcup_{k\in\mathcal{M}}\Psi_{R,k}$  \\ \hline 
            $\rho_k$ & The probability that a randomly selected RIS has $k$ meta-surfaces  \\ \hline  
                $\mu$ & The fraction of blockages that are equipped with RISs.  \\ \hline    
    $\lambda_R$; $\lambda_{R,k}$ & The density of $\Psi_R$; $\Psi_{R,k}$. Note that $\lambda_R=\mu\lambda_b$ and $\lambda_{R,k}=\rho_k\lambda_R$  \\ \hline             
         $P_{\rm LoS}(r)$ & The probability that a user-BS link with length $r$ is clear of any blockages \\ \hline 
         $P_{\rm NLoS}(r)$ & The probability that the link is obstructed by at least one blockage where $P_{\rm NLoS}(r)=1-P_{\rm LoS}(r)$ \\ \hline 
         $\mathcal{E}$; $P_{v}(r)$ & The average ratio of blind-spot areas; the visibility probability for a user-BS pair with distance $r$ apart \\ \hline    
         $\mathcal{A}_i$ & The probability that a randomly selected user is served through an indirect LoS link using an RIS \\ \hline  
    \end{tabular}}
\end{center}
\label{tab:TableOfNotations}
\end{table}
As explained in the introduction, theoretically, the RIS is capable to control the reflected angle of any incident signal~\cite{basar2019wireless}. Hence, the RIS is capable to provide an indirect LoS-link between any transmitter and any receiver as long as both have an LoS with the RIS. As stated earlier, we assume that the RIS is added to one side of the blockage. Hence, this specific RIS-equipped side of the blockage needs to have LoS with a transmitter and a receiver in order to provide them with an indirect LoS-link. Note that a transmitter-receiver pair can have (i) only a direct LoS-link, (ii) only an indirect LoS-link through an RIS, or (iii) both. The case of (iii) can be exploited to enhance the signal quality at the receiver. However, in this paper we focus on a specific use case of the RIS, which is providing indirect LoS-links for transmitter-receiver pairs with no direct LoS-links, as shown in Fig.~\ref{fig:motivation}. This particular use case is of high importance due to its ability to increase the coverage area of any given BS. Hence, large-scale deployment of RISs should eventually lead to reducing the {\em blind-spot} areas of the network. Reducing the blind-spot areas can highly increase the overall coverage probability of the cellular network, specially in the high-frequency bands, which are more sensitive to blockages and require LoS for communication.

To this end, our main purpose throughout this paper is to analyze the considered setup in terms of three specific metrics:
\begin{itemize}
\item The average ratio of blind-spot areas: $\mathcal{E}$.
\item The probability that a given user is served through an indirect link: $\mathcal{A}_i$.
\item The probability that the path-loss between the typical user and its associated BS is below a predefined threshold: $P_{\rm cov}$.
\end{itemize}
We aim to derive each of the above performance metrics as a function of $\lambda_b$ and $\mu$ in order to provide insightful expressions that can be useful for the large-scale deployment of RISs in different kinds of environments. 

\begin{figure}
\centering
\includegraphics[width=0.7\columnwidth]{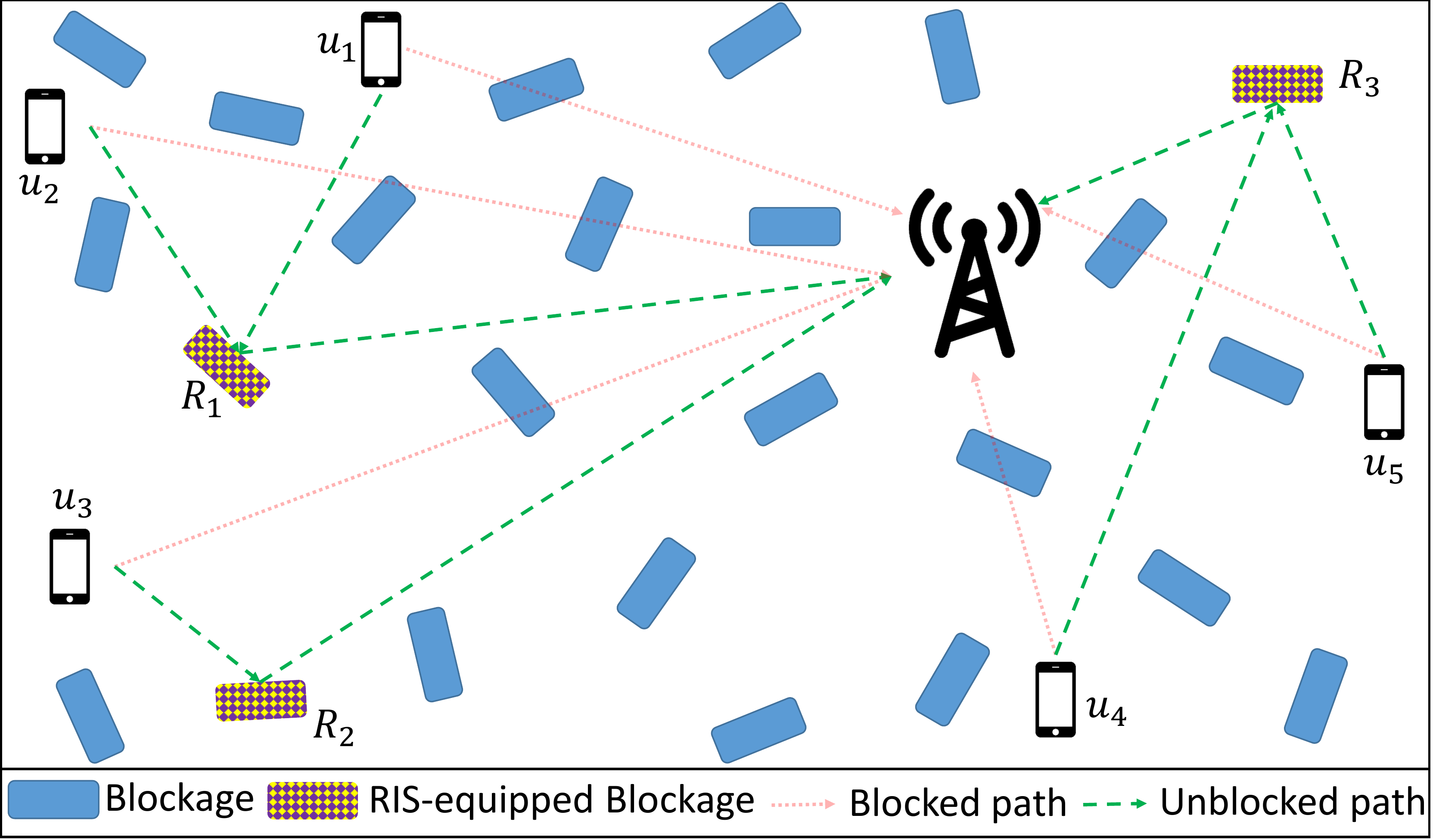}
\caption{ An example of equipping blockages with RISs to provide indirect LoS links for mobile users with blocked paths to the BS.}
\label{fig:planned}
\end{figure}

Before proceeding, it is worth mentioning that in this paper the deployment of the RISs is assumed to be random. In other words, there is no underlying criteria in the selection of locations of the blockages that will be equipped with RISs. Hence, the derived performance metrics throughout the paper provide a lower bound for the performance of the system when the deployment of the RISs is well-planned. A more efficient deployment of RISs can be achieved by selecting the blockages with strategic locations. In other words, blockages that have LoS with blind-spot areas are the best candidates for RIS deployment. In Fig.~\ref{fig:planned}, we show an example of a planned RIS deployment where the selected blockages for RIS deployment are carefully chosen based on their ability to provide indirect LoS-links to the largest possible area.

\subsection{Path-Loss Model}\label{subsec:PL}
In order to evaluate the efficiency of large-scale deployment of RISs, we need to select an appropriate model for the path-loss experienced by the signal received through an indirect path using an RIS. Authors in~\cite{ntontin2019reconfigurable} have provided a detailed discussion about how the received signal power scales with the distance traveled through the indirect path. In particular, the path-loss model highly depends on the size of the meta-surfaces at the RIS relative to the wave-length of the signals. In this paper, we consider the scenario of high-frequency bands where the size of the meta-surfaces is considerably larger than the wavelength. Hence, according to~\cite{ntontin2019reconfigurable}, the power of the received signal through an indirect path scales with $M^2(d_{U-I}+d_{I-B})^{-\alpha}$, where $d_{U-I}$ is the distance between the user and the RIS, $d_{I-B}$ is the distance between the RIS and the BS, $M$ is the number of meta-surfaces in the RIS, and $\alpha$ is the path-loss exponent. On the other hand, in the case of direct LoS-link between the user and the BS, the received signal through the direct path scales with $d_{U-B}^{-\alpha}$, where $d_{U-B}$ is the distance between the user and the BS. 
\subsection{Number of Meta-Surfaces per RIS}
In this paper, we consider two scenarios for the value of $M$. The first scenario is assuming that $M=M_F$ is fixed for all the RISs. The second scenario is assuming that $M$ is a random variable with probability mass function (PMF) $\mathbb{P}(M=k)=\rho_k$ for $k\in\mathcal{M}$, where $\mathcal{M}$ is the set of possible values for $M$. Given that the first scenario is a special case of the second scenario, we will provide our results mainly for the second scenario while highlighting how the final expressions would change when the value of $M$ is fixed. Hence, $\Psi_R$ can be viewed as the superposition of the set of PPPs $\Psi_{R,k}$  $\left(\Psi_R=\bigcup_{k\in\mathcal{M}}\Psi_{R,k}\right)$, where $\Psi_{R,k}$ represents the locations of RISs that have $k$ meta surfaces, and has density $\lambda_{R,k}=\rho_k\lambda_R=\rho_k\mu\lambda_b$. For a fair comparison between the two scenarios, whenever needed, we will make sure that the compared scenarios have the same average number of meta surfaces per unit area: $M_F\lambda_R=\mathbb{E}[M]\lambda_R$.
\subsection{Association Policy}\label{sec:asso}
Following the discussion in Sec.~\ref{subsec:PL}, we consider an association policy that is based on the average path-loss. In particular, the user associates with the BS that provides the {\em lowest average path-loss} (either through  direct or indirect LoS-link). To provide a formal definition for the association policy, we first define the BSs located at $y^d$ and $y^{i,k}$ that provide the lowest path-loss through direct LoS link, and the lowest path-loss through an indirect LoS-link using a RIS with $k$ meta-surfaces, respectively. They can be formally defined as follows
\begin{align}\label{yJ1}
y^d={\rm arg}\underset{y\in\Psi_{BS}}{\rm min}{\rm PL}^d(y),
\end{align}
\begin{align}\label{yJ2}
y^{i,k}={\rm arg}\underset{y\in\Psi_{BS}}{\rm min}{\rm PL}^{i,k}(y),
\end{align}
where 
\begin{align}\label{eq:ld}
{\rm PL}^d(y)=\frac{1}{\delta_{\rm L}^{u_o,y}\|y-u_o\|^{-\alpha}},
\end{align}
\begin{align}\label{eq:li}
{\rm PL}^{i,k}(y)=\frac{1}{k^2\left(1-\delta_{\rm L}^{u_o,y}\right)\underset{z\in\Psi_{R,k}}{\max\ }\delta_{\rm L}^{u_o,z}\delta_{\rm L}^{y,z}\kappa^{u_o,y}_z(\|u_o-z\|+\|y-z\|)^{-\alpha}},
\end{align}
$u_o$ is the location of the typical user, $\delta_{\rm L}^{a,b}=1$ if the path between the locations $a$ and $b$ is free of blockages, and $\delta_{\rm L}^{a,b}=0$ otherwise. The value of $\kappa^{u_o,y}_z=1$ if the RIS whose midpoint is located at $z$ has an orientation that enables it to provide an indirect path between the user located at $u_o$ and the BS located at $y$, otherwise, $\kappa^{u_o,y}_z=0$. The RIS-equipped blockage located at $z$ can provide an indirect path between $u_o$ and $y$ as long as both $u_o$ and $y$ are facing the same side (the RIS-equipped side) of the blockage. 
 Now, defining ${\rm PL}(y)=\min\left({\rm PL}^d(y),\underset{k\in\mathcal{M}}{\rm min}{\rm PL}^{i,k}(y)\right)$,  the association policy can be defined as follows
\begin{align}
y^*={\rm arg}\underset{y\in\Psi_{BS}}{\rm min}{\rm PL}(y).
\end{align}
Note that we are assuming that RISs are solely deployed to support the non-line-of-sight (NLoS) links. Hence, if there exists a direct LoS between the user and the BS, RISs are not used to provide indirect LoS for this specific link.

\subsection{Performance Metrics}
\begin{figure}%
 \centering
 \subfloat[]{\includegraphics[width=0.4\columnwidth]{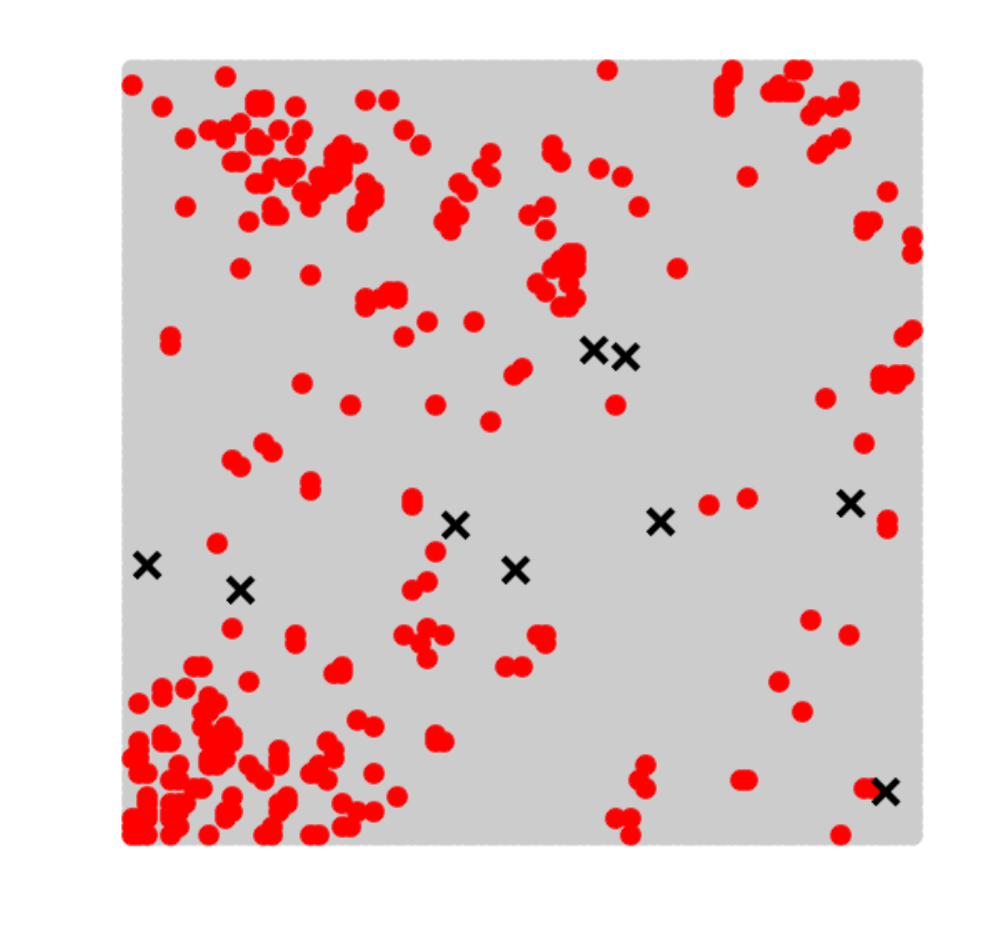}\label{fig3:a}}%
 \subfloat[]{\includegraphics[width=0.4\columnwidth]{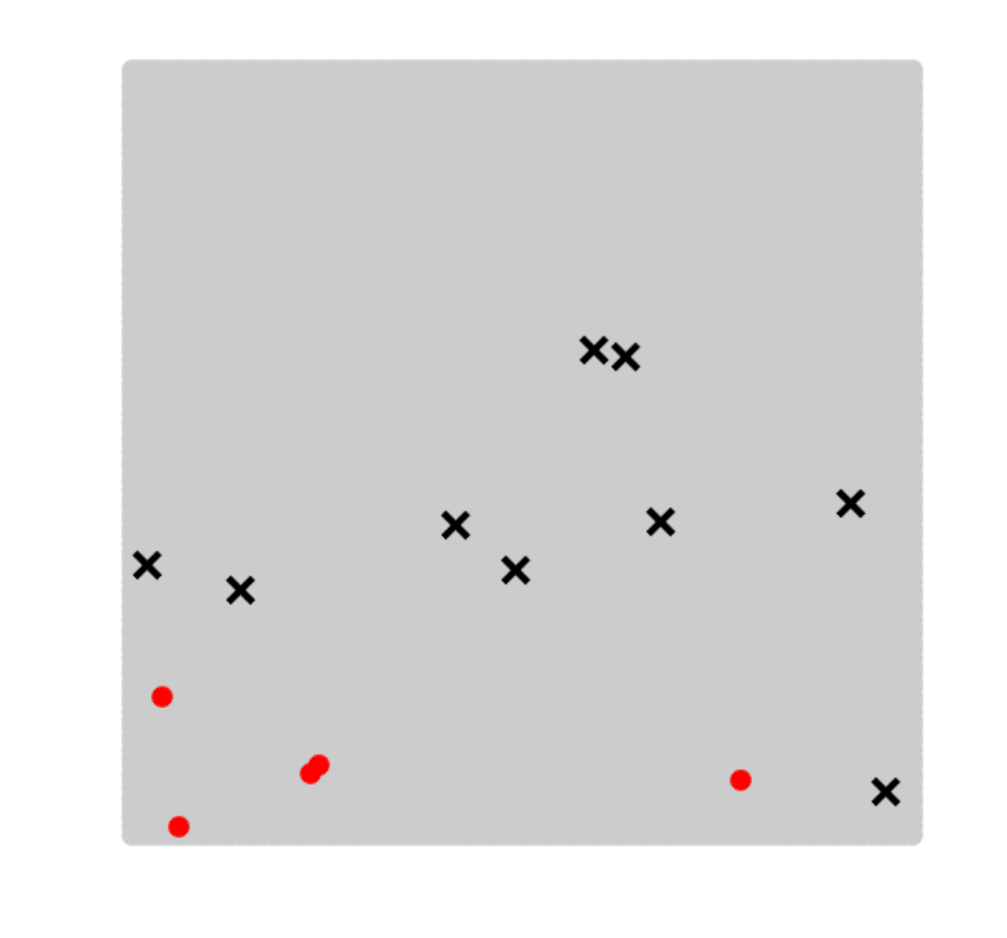}\label{fig3:b}}\\
 \subfloat[]{\includegraphics[width=0.4\columnwidth]{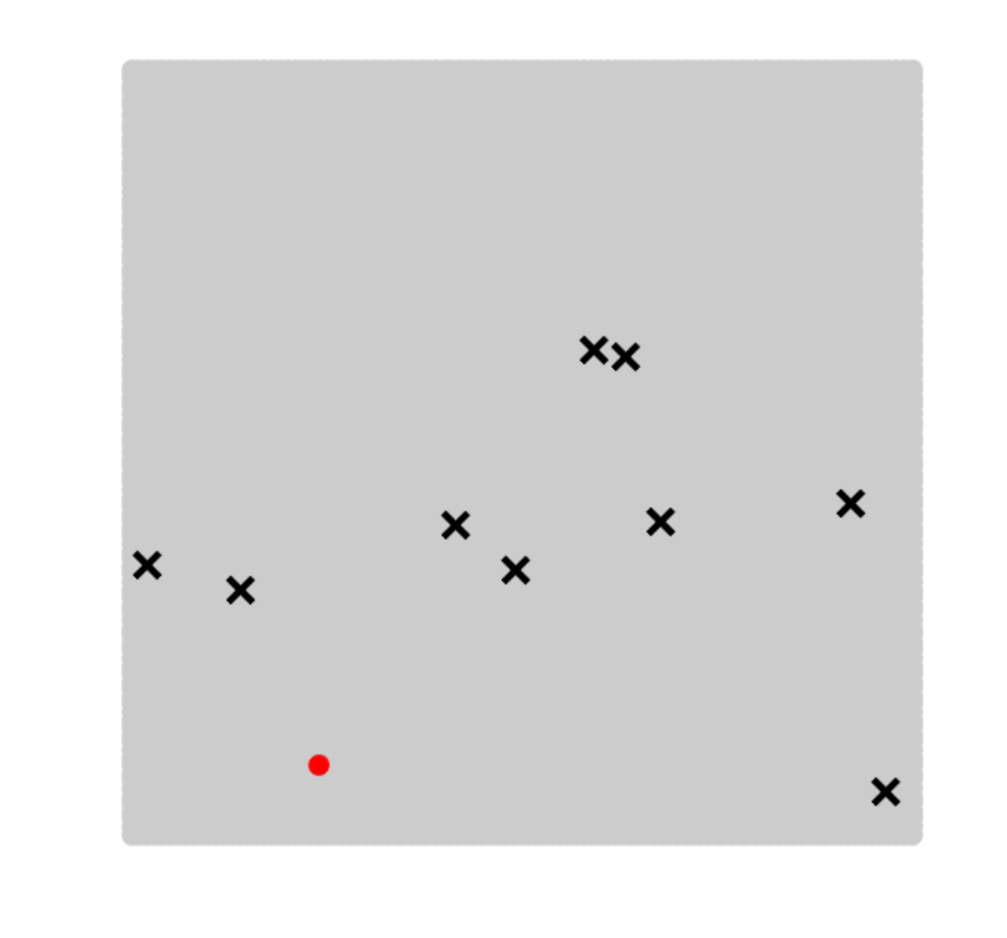}\label{fig3:c}}%
  \subfloat[]{\includegraphics[width=0.4\columnwidth]{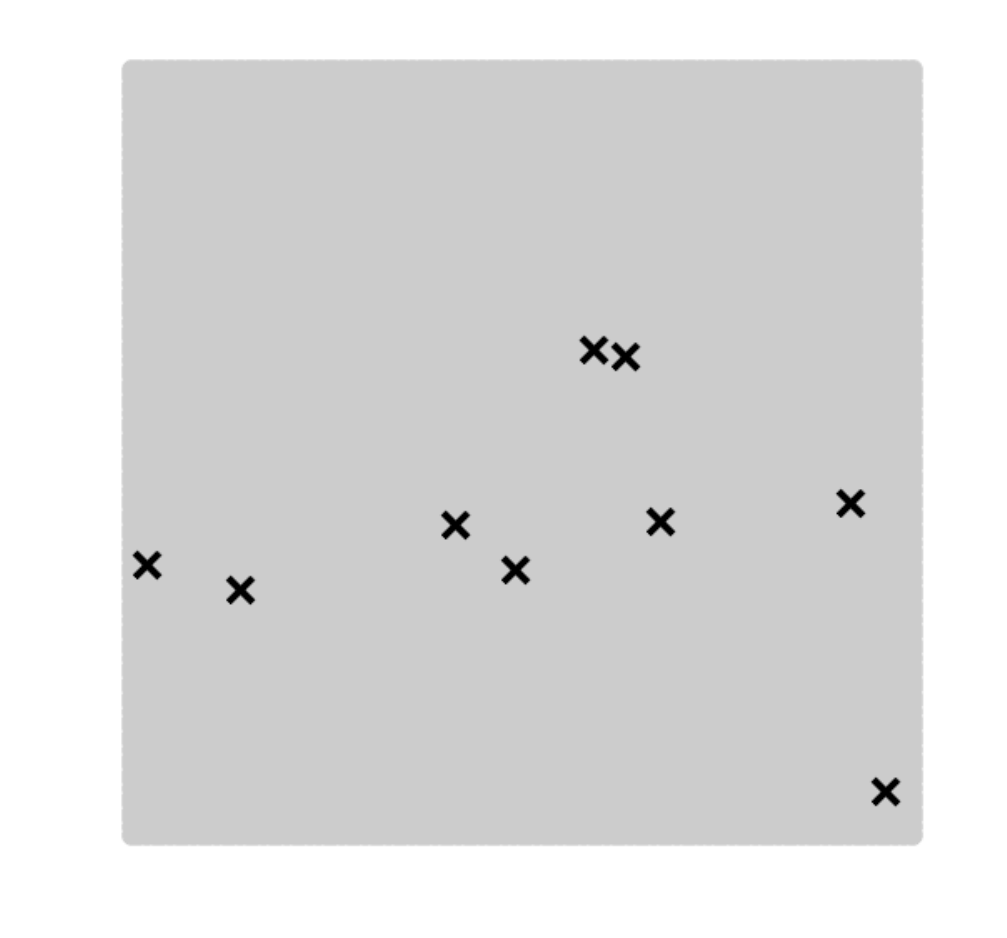}\label{fig3:d}}%
 \caption{A realization of the considered setup with the density of blockages $\lambda_b=500$ blockages/km$^2$, the density of the RISs is $\lambda_R=\mu\lambda_b$, the density of BSs is $10$ BSs/km$^2$, and the simulated area is a square of $1$ km side length. The red spots represent the blind-spot areas, the grey areas are the locations that have at least one LoS-link (either direct or indirect), and the markers represent the locations of the BSs in this realization. The value of $\mu$ varies as follows: (a) $\mu=0$, (b) $\mu=0.05$, (c) $\mu=0.1$, and (d) $\mu=0.4$.}%
 \label{fig:realization}%
\end{figure}

Now, we formally define the main performance metrics that will be derived in Sec.~\ref{sec:analy}. The first performance metric of interest is the average ratio of blind-spot areas, which is described in Definition~\ref{def:not}. This metric can be formally defined as follows
\begin{align}\label{not:def_eq}
\mathcal{E}=\mathbb{E}\left[\prod_{y\in\Psi_{BS}}\left(1-\left(\delta_{\rm L}^{u_o,y}+(1-\delta_{\rm L}^{u_o,y})\left(1-\prod_{z\in\Psi_R}\left(1-\delta_{\rm L}^{z,u_o}\delta_{\rm L}^{z,y}\kappa_z^{u_o,y}\right)\right)\right)\right)\right].
\end{align}
Note that the product inside the expectation equals to 1 only if there is neither direct nor indirect LoS-links to any BS. In Fig.~\ref{fig:realization}, we show a realization of the considered setup and how equipping just $5\%$ of the blockages with RISs highly reduces the area of blind-spots. We observe that when the value of $\mu=0.4$ ($40\%$ of the blockages are equipped with RISs) the blind-spot areas completely disappear.

The next metric of interest is the efficiency of the RIS deployment. This can be evaluated by deriving the probability of association through an RIS $\mathcal{A}_i$. Note that a user can be in one of three states: (i) associated with a BS through a direct link, (ii) associated with a BS through an indirect link, and (iii) falling in a blind-spot. The value of $\mathcal{A}_i$ can be used to estimate the ratio of utilized RISs as will be explained in detail in Remark~\ref{rem4}. The association probability is defined as follows
\begin{align}
\mathcal{A}_i=1-\mathbb{P}\left(\bigcap_{k\in\mathcal{M}}{\rm PL}^d\left(y^d\right)\leq {\rm PL}^{i,k}\left(y^{i,k}\right)\right)-\mathcal{E},
\end{align} 
where $y^d$, $y^{i,k}$, ${\rm PL}^d(y)$, and ${\rm PL}^{i,k}(y)$ are defined in (\ref{yJ1}), (\ref{yJ2}), (\ref{eq:ld}), and (\ref{eq:li}), respectively. 

Finally, we define the coverage probability as the probability that the path-loss ${\rm PL}(y^*)$ is below a predefined threshold $\tau$:
\begin{align}\label{cov:def}
P_{\rm cov}=\mathbb{P}\left({\rm PL}(y^*)\leq\tau\right).
\end{align}

Before deriving each of the performance metrics introduced in this section, we first provide some preliminaries in the next section, which will make it easier to follow the mathematical derivations provided throughout the paper.
\section{Preliminaries on Stochastic Geometry}
Tools from stochastic geometry has been widely used in literature during the past decade~\cite{7733098}. The majority of the literature focuses on utilizing homogeneous PPP for modeling the locations of wireless network components. On the other hand, inhomogeneous PPP is rarely used in literature. Since majority of the analysis in this paper relies on using inhomogeneous PPP, we provide in this section some preliminaries that will facilitate understanding the mathematical derivations provided in the rest of this paper. 

For any homogeneous PPP $\Phi_h$ in $\mathbb{R}^2$ with density $\lambda_h$, the number of points in any given area $B\subset\mathbb{R}^2$ is a Poisson distributed random variable with mean $\lambda_h |B|$. On the other hand, for an inhomogeneous PPP $\Phi_i$ with density $\lambda_i(x)$, the number of points in $B\subset \mathbb{R}^2$ is a Poisson distributed random variable with mean $\int_B \lambda_i(x) {\rm d}x$.

One of the key important characteristics of PPPs is the void probability, which is the probability of having zero points within a given area $B$. In particular, it is typically used to compute the contact distance distribution, which is the distribution of the distance between the origin and the nearest point in the PPP. Based on the above discussion, the void probability of a homogeneous PPP in $B$ is $\exp(-\lambda_h |B|)$, while for the inhomogeneous PPP it is $\exp\left(-\int_B\lambda_i(x){\rm d}x\right)$. Next, the distribution of the contact distance $R_c$ can be computed for homogeneous PPPs as follows:
$$F_{R_c}(y)=1-\mathbb{P}(\mathcal{N}_{\Phi_h}(\mathcal{B}(0,y))=0)=1-\exp(-\lambda_h \pi y^2), $$
while for inhomogeneous PPPs it can be computed as follows
$$F_{R_c}(y)=1-\mathbb{P}(\mathcal{N}_{\Phi_i}(\mathcal{B}(0,y))=0)=1-\exp\left(-\int_{\mathcal{B}(0,y)}\lambda_i(x){\rm d}x\right), $$
where $\mathcal{B}(0,y)$ is a disk centered at the origin with radius $y$, $\mathcal{N}_{\Phi_h}(B)$ and $\mathcal{N}_{\Phi_i}(B)$ are the number of points in $B$ for homogeneous and inhomogeneous PPPs, respectively.

Thinning is typically used in literature to refer to the process of removing points from a point process according to a specific probability distribution. The thinning process is described as {\em independent thinning} if the probability of removing each point is independent from the other points. One category that falls under this description is {\em location-dependent} thinning, which is defined next.
\begin{definition}[Location-Dependent Thinning]
For a given point process $\Phi$, location-dependent thinning is achieved by removing each point $x\in\Phi$ with probability $1-g(x)$, where $g(x)$ is only function of the location of the point, and independent of the locations of the other points in $\Phi$.
\end{definition}

In the next proposition, we provide a result that will be frequently used throughout this paper, which is related to location-dependent thinning of homogeneous PPPs.
\begin{prop}[Location-Dependent Thinning of homogeneous PPP]\label{prelim:lem1}
If $\Phi$ is a homogeneous PPP with density $\lambda$, the location-dependent thinning of $\Phi$ with probability $1-g(x)$ leads to an inhomogeneous PPP $\tilde{\Phi}$ with density $\tilde{\lambda}(x)=g(x)\lambda$.
\begin{IEEEproof}
The proof is provided in~\cite[Theorem 2.36]{haenggi2012stochastic} and is hence skipped.
\end{IEEEproof}
\end{prop}
Tractable analysis of inhomogeneous PPPs can be achieved, specially when its density function $\tilde{\lambda}(x)$ has some properties. We define one important property next.
\begin{definition}[Radially-Symmetric Intensity]
An inhomogeneous PPP $\tilde{\Phi}$ has a radially-symmetric intensity if its density $\tilde{\lambda}(x)$ is only function of $\|x\|$ (i.e., the distance between $x$ and the origin).
\end{definition}
\begin{definition}[Isotropic Point Process]
A point process is isotropic if it is rotation-invariant when rotated around the origin.
\end{definition}
Using the above two definitions, we are now ready to provide an important proposition, which will be used frequently in the analytical part of this paper.
\begin{prop}[Isotropic Inhomogeneous PPP]\label{prop2}
An inhomogeneous PPP with radially-symmetric density is isotropic~\cite{haenggi2012stochastic}.
\end{prop}
Now that we have enlisted the key preliminaries required for grasping the mathematical proofs in the analytical part of the paper, we move forward to study the performance of the considered system in the next section.

\section{Performance Analysis}\label{sec:analy}
This section contains the main technical contributions of this paper. In particular, we provide mathematical expressions for the performance metrics formally defined in Sec.~\ref{sec:sys}. In the rest of the paper, without loss of generality,  we assume that the location of the typical user $u_o$ is the origin, and hence will be dropped from all the notations for simplicity. For the case of uniformly distributed blockage orientation with length $L$ and density $\lambda_{b}$, the LoS probability between a user and a BS at a distance $r$ was derived in~\cite{6840343} as follows:
\begin{align}
P_{\rm LoS}(r)=\exp\left(-\beta{r}\right),
\end{align} 
where $\beta=\frac{2\lambda_b \mathbb{E}[L]}{\pi}$. This leads to the cumulative distribution function of the distance $R_d$ between the typical user and its nearest LoS BS as follows
\begin{align}\label{eq:FRd}
F_{R_d}(x)=1-\exp\left(-2\pi\lambda_{BS}\frac{1}{\beta^2}\Big(1-(\beta x+1)\exp(-\beta x)\Big)\right),
\end{align}
which can be derived by using $P_{\rm LoS}(r)$ as a thinning probability for $\Psi_{BS}$, applying Proposition~\ref{prelim:lem1}, and then computing the null probability for the resulting inhomogeneous PPP. The probability density function of $R_d$ can be then derived as follows
\begin{align}
f_{R_d}(x)=2\pi\lambda_{BS}x\exp\left(-\beta x-2\pi\lambda_{BS}\frac{1}{\beta^2}\Big(1-(\beta x+1)\exp(-\beta x)\Big)\right).
\end{align}
\subsection{blind-spot Areas}
Our first objective is to derive the fraction of blind-spot areas $\mathcal{E}$, provided in Definition~\ref{def:not}. We first simplify the mathematical definition of $\mathcal{E}$ provided in (\ref{not:def_eq}) as follows

\begin{align}\label{not:def:fin}
\mathcal{E}&=\mathbb{E}\Bigg[\prod_{y\in\Psi_{BS}}\Bigg(1-\Bigg(\delta_{\rm L}^{y}+(1-\delta_{\rm L}^{y})\left(1-\prod_{z\in\Psi_R}\left(1-\delta_{\rm L}^{z}\delta_{\rm L}^{z,y}\kappa_z^{y}\right)\right)\Bigg)\Bigg)\Bigg]\nonumber\\
&\overset{(a)}{=}\mathbb{E}_{\Psi_{BS}}\Bigg[\prod_{y\in\Psi_{BS}}\Bigg(1-\Bigg(P_{\rm LoS}(\|y\|)+P_{\rm NLoS}(\|y\|)\times\nonumber\\& \ \ \ \ \ \ \ \ \ \ \ \ \ \ \ \ \ \ \ \ \ \ \ \ \ \ \ \  \ \ \ \ \mathbb{E}_{\Psi_R}\left[1-\prod_{z\in\Psi_R}\left(1-a_i(\|y\|,\|z\|,\phi)\right)\right]\Bigg)\Bigg)\Bigg]\nonumber\\
&\overset{(b)}{=}\exp\Bigg(-2\pi\lambda_{BS}\int_0^\infty \Bigg(P_{\rm LoS}(r)+P_{\rm NLoS}(r)\underbrace{\mathbb{E}_{\Psi_R}\left[1-\prod_{z\in\Psi_R}\left(1-a_i(r,\|z\|,\phi)\right)\right]}_{P_I(r)}\Bigg)r{\rm d}r\Bigg),
\end{align}
where $(a)$ comes from assuming that the number of blockages experienced by different links in the network are independent, $(b)$ follows by using the probability generating functional (PGFL) of PPP. The value $a_i(r,t,\phi)$ represents the probability that an RIS at distance $t$ from the origin can provide an indirect LoS path between the typical user and a BS at distance $r$ from the origin, with $\phi$ being the angle between the user-RIS and user-BS links. The value of $P_I(r)$ represents the probability of having at least one RIS capable of providing an indirect LoS path between the typical user and a BS at a distance $r$ from the origin. In the following lemma, we derive the probability that a given RIS satisfies all the conditions required for it to be able to provide an indirect path for a given user-BS link.
\begin{figure}
\centering
\includegraphics[width=0.6\columnwidth]{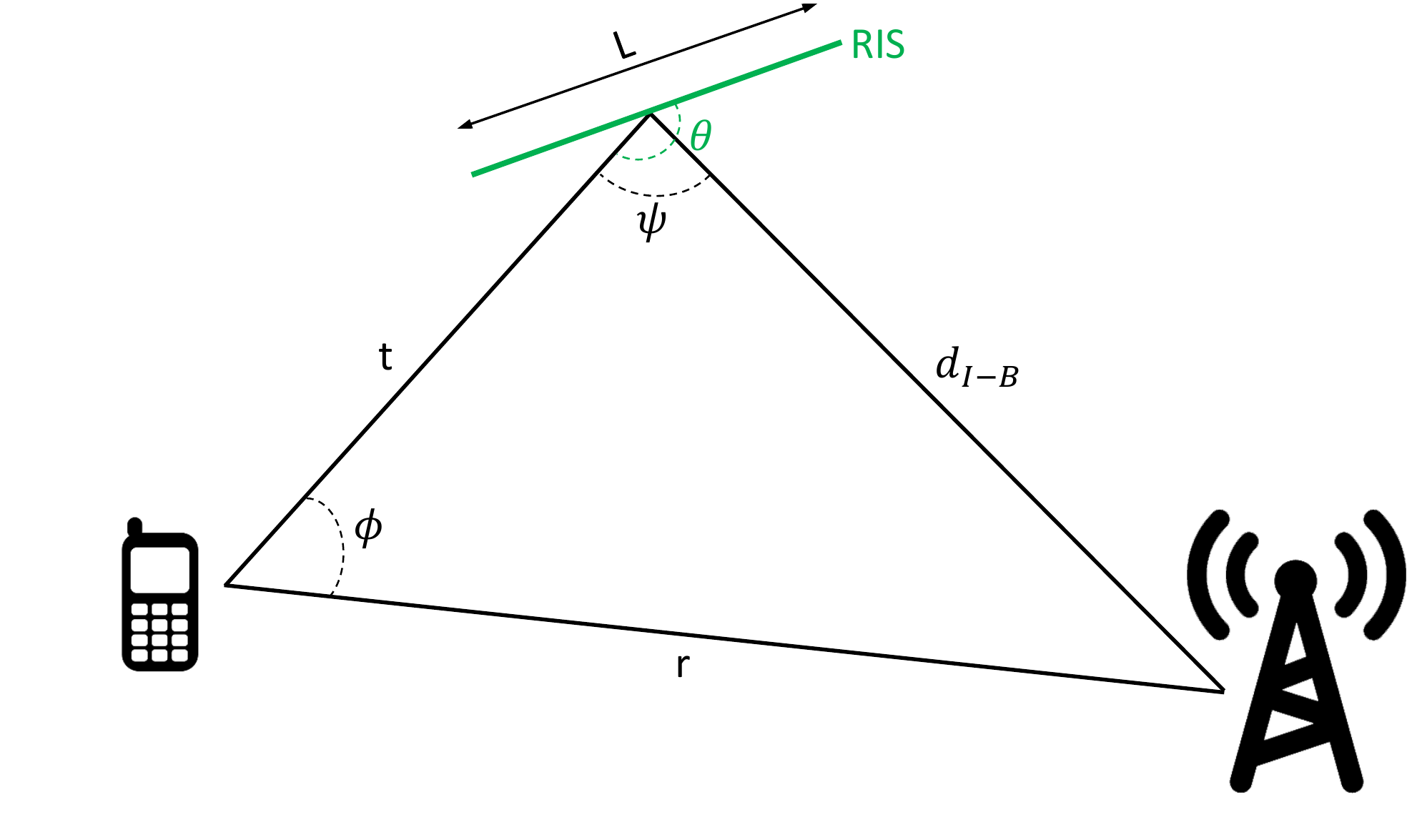}
\caption{The setup for Lemma~\ref{lemma1}.}
\label{fig:proof1}
\end{figure}
\begin{lemma}[Probability of Reflection]\label{lemma1}
For the setup shown in Fig.~\ref{fig:proof1}, with a distance $t$ between the typical user and the RIS, distance $r$ between the user and the BS, and angle $\phi$ between the user-BS link and the user-RIS link, the probability that the RIS is capable of providing an indirect LoS path for the user-BS link is
\begin{align}
a_i(r,t,\phi)=\frac{1}{2}P_{\rm LoS}(t)P_{\rm LoS}\left(\sqrt{r^2+t^2-2rt\cos(\phi)}\right)\mathcal{C}(r,t,\phi),
\end{align}
where $\mathcal{C}(r,t,\phi)=1-\frac{1}{\pi}\cos^{-1}\left(\frac{t-r\cos(\phi)}{\sqrt{t^2+r^2-2rt\cos(\phi)}}\right)$.
\begin{IEEEproof}
See Appendix~\ref{app:lemma1}.
\end{IEEEproof}
\end{lemma}
\begin{remark}\label{rem1}
The expression in Lemma~\ref{lemma1} represents the joint probability of satisfying the three conditions: (i) having an LoS with the user, (ii) having an LoS with the BS, and (iii) having an orientation that enables the indirect path. One interesting insight can be drawn from the expression in Lemma~\ref{lemma1} is the effect of $t$ on the value of $a_i(r,t,\phi)$. We observe that as $t$ increases, for a given $r$, the effect of the orientation of the RIS reduces. This can be observed from the fact that $\mathcal{C}(r,t,\phi)$ increases and approaches unity as $t$ increases. However, on the other hand, increasing $t$ reduces the probability of having an LoS-link between the RIS and either the user or the BS. This can be observed from the fact that each of $P_{\rm LoS}(t)$ and $P_{\rm LoS}\left(\sqrt{r^2+t^2-2rt\cos(\phi)}\right)$ approach zero as $t$ increases. Hence, there is a trade-off in the effect of the value of $t$ on $a_i(r,t,\phi)$. 
\end{remark}
The above remark provides an important design insight for deployment of RISs. In particular, it implies that RISs serving a specific area might not be the closest RISs to this area. It is worth reminding that the drawn insights are specific to the setup considered in this paper where the deployment of the RISs is random and unplanned.

Recalling Proposition~\ref{prelim:lem1}, the result in Lemma~\ref{lemma1} can be used as a location-dependent thinning probability for the PPP $\Psi_R$. Hence, the resulting inhomogeneous PPP represents the locations of RISs that are capable of providing an indirect path between the typical user and a BS at a distance $r$. In the following lemma, we derive the probability of having at least one RIS that is capable of providing an indirect path between the typical user and a BS at a distance $r$.
\begin{lemma}[Indirect Path Probability]\label{lemma2}
The probability of having an indirect link, through an RIS, between the typical user and a BS at distance $r$ is
\begin{align}
P_I(r)=1-\exp\left(-\lambda_{R}\int_{-\pi}^{\pi}\int_0^\infty a_i(r,t,\phi)t{\rm d}t{\rm d}\phi\right).
\end{align}
\begin{IEEEproof}
See Appendix~\ref{app:lemma2}.
\end{IEEEproof}
\end{lemma}
With the existence of RISs, the set of {\em visible} BSs by the typical user are split into two categories: (i) BSs with direct LoS paths from the user, and (ii) BSs which are visible through at least one indirect path. A BS is visible by the typical user through indirect path if there exists at least one RIS that has an LoS-link with both the typical user and the BS. The density of visible BSs is provided next.
\begin{lemma}[Density of Visible BSs]\label{lemma3}
For the typical user, located at the origin, the density of visible BSs through either direct or indirect paths is
\begin{align}
\lambda_v(r)=\lambda_{BS}P_v(r),
\end{align}
where $P_v(r)$ is the {\em visibility probability}, defined by the probability that a BS at distance $r$ from the typical user is visible through either direct or indirect paths, and $P_v(r)=P_{\rm LoS}(r)+P_{\rm NLoS}(r)P_I(r)$.
\begin{IEEEproof}
As stated above, a BS at distance $r$ from the typical user is visible if there exists either a direct LoS-link or at least one indirect LoS-link. The probability of having a direct LoS-link is $P_{\rm LoS}(r)$. On the other hand, the probability of not having a direct link but at least one indirect link is $P_{\rm NLoS}(r)P_I(r)$. Hence, given the two events are mutually exclusive, the probability of the BS being visible by the typical user is $P_v(r)=P_{\rm LoS}(r)+P_{\rm NLoS}(r)P_I(r)$. Hence, the locations of visible BSs are modeled by an inhomogeneous PPP $\Psi_v$ with density $\lambda_{BS}P_v(r)$.
\end{IEEEproof}
\end{lemma}
\begin{remark}\label{rem2}
The gained coverage enhancement from the deployment of RISs can be observed in Lemma~\ref{lemma3}. Before the deployment of RISs, the density of visible BSs is $\lambda_{BS}P_{\rm LoS}(r)$. After adding RISs, the density of visible BSs increases by $\lambda_{BS}P_{\rm NLoS}(r)P_I(r)$. The value of the density of RISs appears implicitly in $P_I(r)$ as we recall from Lemma~\ref{lemma2}.
\end{remark}

Now we are ready to provide one of the main results of this section, the fraction of blind-spot areas, in the below theorem. 
\begin{theorem}[Fraction of Blind-Spots]\label{thm1}
The fraction of blind-spot areas is
\begin{align}
\mathcal{E}=\exp\left(-2\pi\int_0^\infty \lambda_v(r)r{\rm d}r\right).
\end{align}
\begin{IEEEproof}
This result follows directly by substituting in (\ref{not:def:fin}) using the result in Lemma~\ref{lemma2}. 
\end{IEEEproof}
\end{theorem}
\begin{remark}\label{rem3}
Similar to the discussion in Remark~\ref{rem2}, we can also evaluate the performance enhancement gained from deploying RISs using the expression in Theorem~\ref{thm1}. Before deploying RISs, the fraction of blind-spots is $\exp\left(-2\pi\int_0^\infty\lambda_{BS}P_{\rm LoS}(r)r{\rm d}r\right)$. Comparing with the result in Theorem~\ref{thm1}, we observe that deploying RISs reduces the fraction of blind-spots by a factor of $\exp\left(-2\pi\int_0^\infty\lambda_{BS}P_{\rm NLoS}(r)P_I(r)r{\rm d}r\right)$. Recalling that $\lambda_R=\mu\lambda_b$, we can observe from Lemma~\ref{lemma2} that $P_I(r)$ is an increasing function of $\mu$. Hence, the value of $\mathcal{E}$, as explained, is a decreasing function of $\mu$.
\end{remark}
In the next part of this section, our objective is to derive the probability distribution of the shortest visible path length. 
\subsection{Indirect LoS-link Length Distribution}
In order to study the association between the users and the BSs either through direct LoS-links or through RISs, we first need to study the statistics of the length of the indirect LoS-links. In the following Lemma, we derive the probability distribution of the shortest visible indirect LoS-link length between the typical user and a BS at distance $r$ through an RIS with $k$ meta-surfaces. 
\begin{lemma}[Indirect Path Length Distribution]\label{lemma4}
The probability distribution of $R_{i,k}|r$, the length of the shortest indirect path between the typical user and a BS at distance $r$ using any RIS with $k$ meta-surfaces, is
\begin{align}\label{cdf_cond1}
F_{R_{i,k}|r}(x)=\left\{
	\begin{array}{ll}
		0  & \mbox{if } {x}< r \\
		1-\exp\left(-\lambda_{R}\rho_k\int_{-\pi}^{\pi}\int_0^{\frac{x^2-r^2}{2(x-r\cos(\phi))}} a_i(r,t,\phi)t{\rm d}t{\rm d}\phi\right) & \mbox{if } {x} \geq r
	\end{array},
\right.
\end{align}
while the probability distribution of $R_{i}|r$, the length of the shortest indirect path between the typical user and a BS at distance $r$ using any RIS, is
\begin{align}\label{cdf_cond2}
F_{R_i|r}(x)=1-\prod_{k\in\mathcal{M}}\bar{F}_{R_{i,k}|r}(x),
\end{align}
where $\bar{F}_{R_{i,k}|r}(x)=1-{F}_{R_{i,k}|r}(x)$.
\begin{IEEEproof}
See Appendix~\ref{app:lemma4}.
\end{IEEEproof}
\end{lemma}

Recalling Proposition~\ref{prelim:lem1}, (\ref{cdf_cond1}) can be used as a thinning probability for the points in $\Psi_{BS}$. The retained points after thinning represent the locations of the BSs whose shortest indirect path through RISs from $\Psi_{R,k}$ has a length less than $x$. The remaining points can be modeled as an inhomogeneous PPP with density $\lambda_{BS}F_{R_{i,k}|r}(x)$. Hence, we can use this result to derive the probability distribution of $R_{i,k}$, which is the shortest indirect path between the typical user and the NLoS BSs through RISs from $\Psi_{R,k}$. Similar approach can be followed to derive the distribution of $R_i$. The distribution of each of $R_{i,k}$ and $R_i$ are provided next.
\begin{lemma}[Probability Distribution of $R_i$ and $R_{i,k}$]\label{lemma:Ri}
The probability distribution of $R_{i,k}$ is
\begin{align}\label{CDF1}
F_{R_{i,k}}(x)=1-\exp\left(-2\pi\lambda_{\rm BS}\int_0^x P_{\rm NLoS}(r)F_{R_{i,k}|r}(x)r{\rm d}r\right),
\end{align}
and the distribution of $R_i$ is
\begin{align}\label{CDF2}
F_{R_{i}}(x)=1-\exp\left(-2\pi\lambda_{\rm BS}\int_0^x P_{\rm NLoS}(r)F_{R_{i}|r}(x)r{\rm d}r\right),
\end{align}
where $F_{R_{i,k}|r}(x)$ and $F_{R_{i}|r}(x)$ are given in Lemma~\ref{lemma4}.
\begin{IEEEproof}
See Appendix~\ref{app:lemma:Ri}.
\end{IEEEproof}
\end{lemma}
Using $F_{R_d}(x)$ and $F_{R_{i}}(x)$, in the following theorem, we derive the CDF of the shortest visible path length, which can be either a direct or an indirect path.
\begin{theorem}\label{thm2}
The CDF of the shortest path-length between the typical user and a visible BS is
\begin{align}
F_W(x)=1-\bar{F}_{R_d}(x)\bar{F}_{R_{i}}(x),
\end{align}
where $\bar{F}_{R_i}(x)=1-F_{R_i}(x)$, $\bar{F}_{R_d}(x)=1-F_{R_d}(x)$, and $F_{R_d}$ is given in (\ref{eq:FRd}).
\begin{IEEEproof}
See Appendix~\ref{app:thm2}.
\end{IEEEproof}
\end{theorem}
\begin{remark}\label{rem:W}
The fraction of blind-spot areas $\mathcal{E}$, derived in Theorem~\ref{thm1}, can also be derived using $F_W(x)$. In particular, $\mathcal{E}=1-F_W(\infty)$. With simple substitutions, we can easily observe that indeed $1-F_W(\infty)$ reduces to the expression in Theorem~\ref{thm1}.
\end{remark}
\subsection{Association Probability and RIS Deployment Efficiency}
A randomly selected user can be in one of the three following states: 
\begin{itemize}
\item Falls in a blind-spot, with probability $\mathcal{E}$,
\item Associated with a BS through a direct-path, with probability $\mathcal{A}_d$,
\item Associated with a BS through an indirect path using an RIS, with probability $\mathcal{A}_i$,
\end{itemize}
where $\mathcal{E}+\mathcal{A}_i+\mathcal{A}_d=1$. While $\mathcal{E}$ is provided in Theorem~\ref{thm1}, we provide expressions for $\mathcal{A}_i$ in the following Theorem.
\begin{theorem}[Association Probability]\label{thm3}
The probability that the typical user is associated with a BS using an RIS is
\begin{align}
\mathcal{A}_i=1-\mathcal{E}-\int_0^\infty f_{R_d}(x)\mathcal{H}(x){\rm d}x,
\end{align}
where $\mathcal{H}(x)=\exp\left(-2\pi\lambda_{BS}\int_0^\infty P_{\rm NLoS}(r)\left(1-\prod_{k\in\mathcal{M}}\bar{F}_{R_{i,k}|r}(xk^\frac{2}{\alpha})\right)r{\rm d}r\right)$.
\begin{IEEEproof}
See Appendix~\ref{app:thm3}.
\end{IEEEproof}
\end{theorem}
\begin{remark}\label{rem4}
Given the density of users $\lambda_u$, the density of users associated with an RIS is $\mathcal{A}_i\lambda_u$. Hence, if $\mathcal{A}_i\lambda_u\gg\lambda_R$, it is ensured that all the deployed RISs are used to provide coverage for users. On the other hand, if $\mathcal{A}_i\lambda_u\leq\lambda_R$, then at least $\frac{\lambda_R-\mathcal{A}_i\lambda_u}{\lambda_R}$ out of the deployed RISs are being unused.
\end{remark}
The above remark provides some useful insights on the efficiency of the RIS deployment and percentage of unused RISs. We define the efficiency of deployment of RISs as $\eta=\min\{1,\frac{\lambda_u\mathcal{A}_i}{\lambda_R}\}$, which is actually an upper bound on the average number of RISs associated with at least one user.
\subsection{Coverage Analysis}
The average path-loss when associating with BS through a direct LoS is ${\rm PL}_d=R_d^{\alpha}$. When the user associates with a BS through an indirect path using an RIS, according to~\cite{ntontin2019reconfigurable}, the average path-loss is ${\rm PL}_i=\left({M}\right)^{-2}R_i^{\alpha}$, where $M$ is the number of meta-surfaces in the RIS. In the following theorem, we derive the coverage probability introduced in (\ref{cov:def}).
\begin{theorem}[Coverage Probability]\label{thm:cov}
The probability that the average path-loss between a typical user and the associated BS is below a predefined threshold $\tau$ is
\begin{align}
P_{\rm cov}=1-\bar{F}_{R_d}\left({\tau}^{\frac{1}{\alpha}}\right)\mathcal{H}\left({\tau}^{\frac{1}{\alpha}}\right),
\end{align}
where $\bar{F}_{R_d}(x)=1-{F}_{R_d}(x)$ and $\mathcal{H}(x)$ is given in Theorem~\ref{thm3}.
\begin{IEEEproof}
See Appendix~\ref{app:thm:cov}.
\end{IEEEproof}
\end{theorem} 

In the next section, we provide numerical results for each of the performance metrics derived in this paper in order to draw useful system-level insights.
\section{Results and Discussion}
In this section we provide the numerical results for the expressions derived throughout the paper. Similar to~\cite{8930608,8941080,8970580}, we use the performance of the system without RISs as the benchmark in order to visualize the gains added by the deployment of RISs. We consider three values for the density of blockages $\lambda_b$: $300$, $500$, and $700$ blockages/km$^2$. The density of the BSs is $\lambda_{BS}=10$ BS/km$^2$, the density of the users is $\lambda_u=300$ user/km$^2$, the average value of $L$ is $15$ m. 
\begin{figure}%
 \centering
 \subfloat[]{\includegraphics[width=0.45\columnwidth]{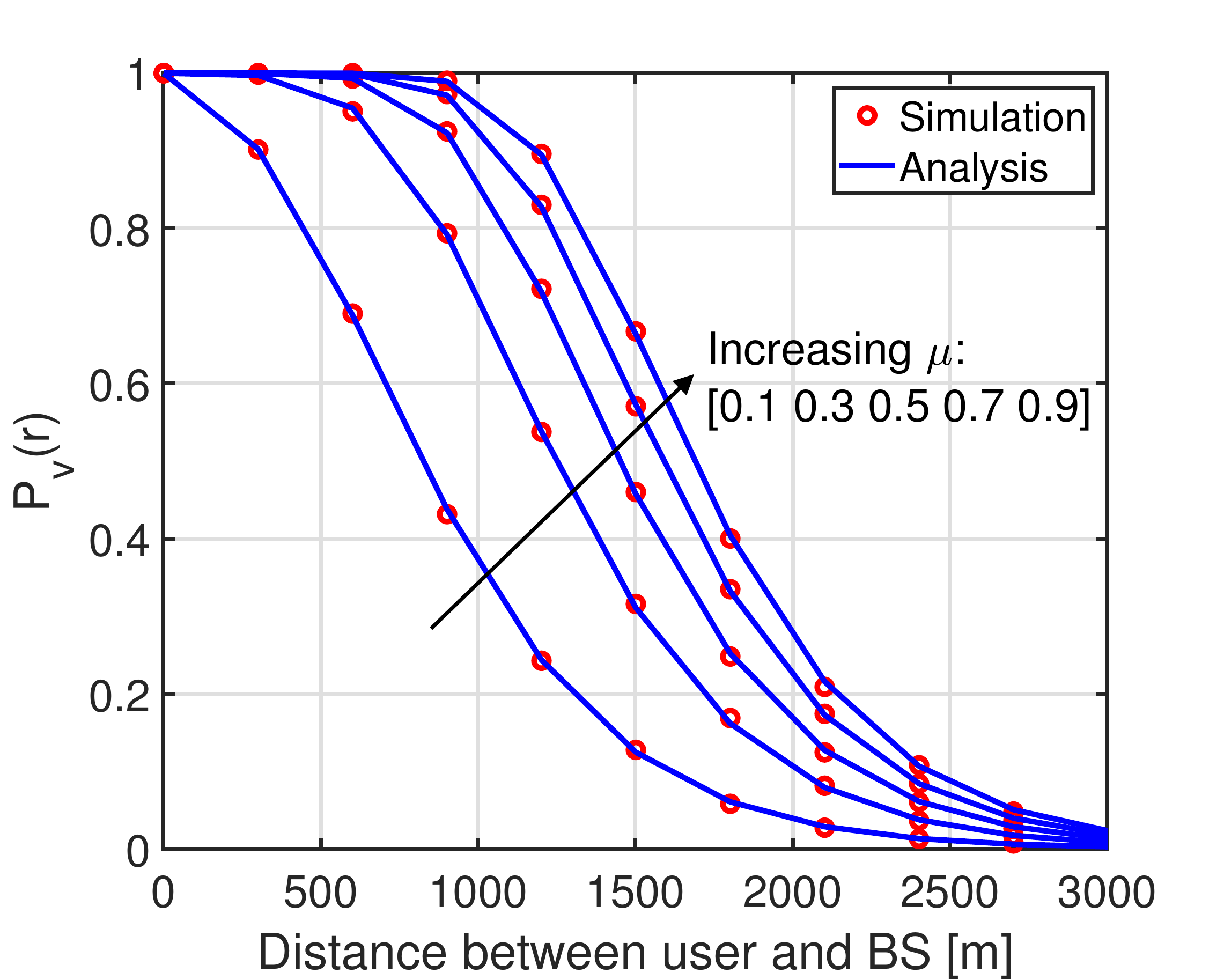}\label{fig1:a}}%
 \subfloat[]{\includegraphics[width=0.45\columnwidth]{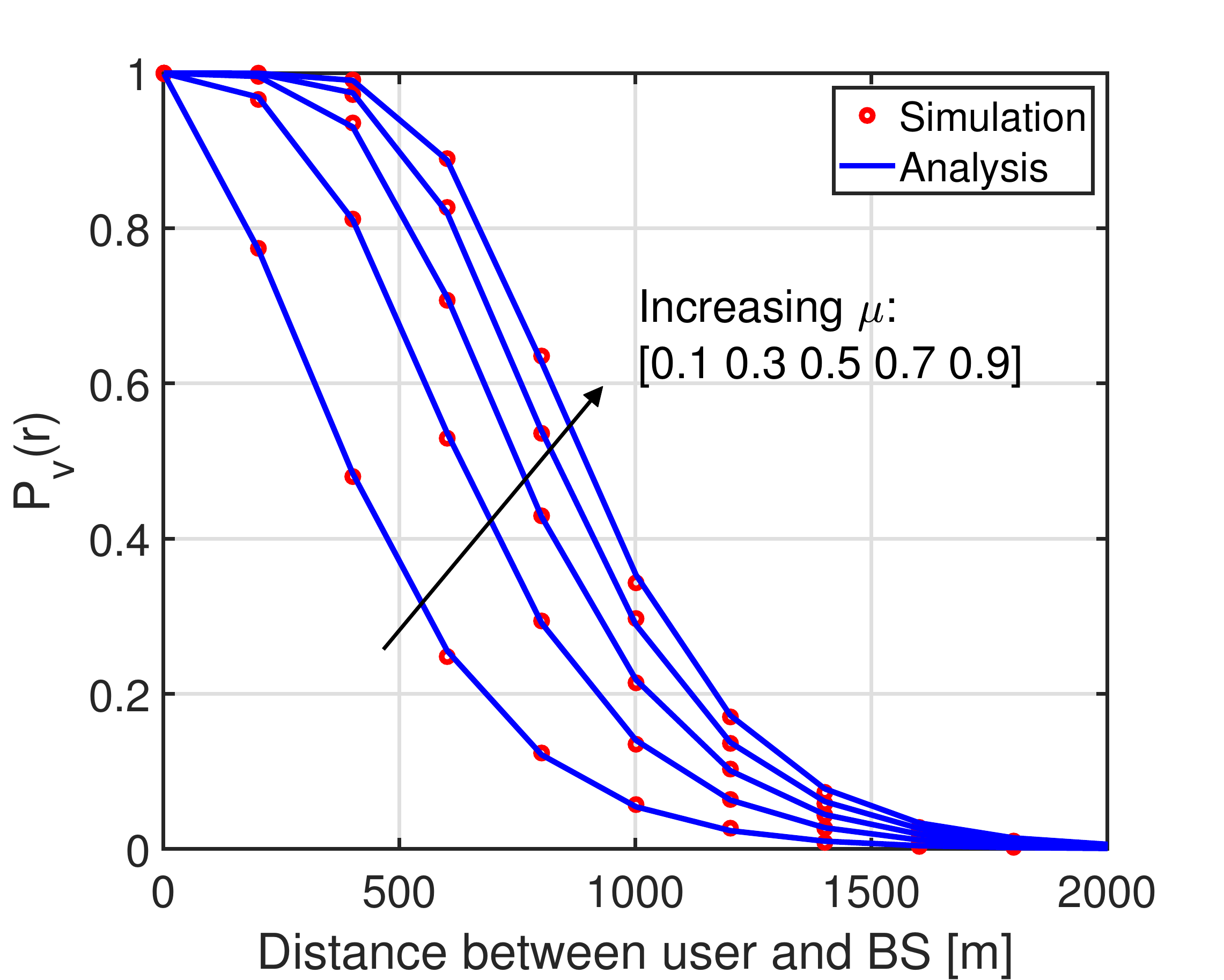}\label{fig1:b}}\\
 \subfloat[]{\includegraphics[width=0.45\columnwidth]{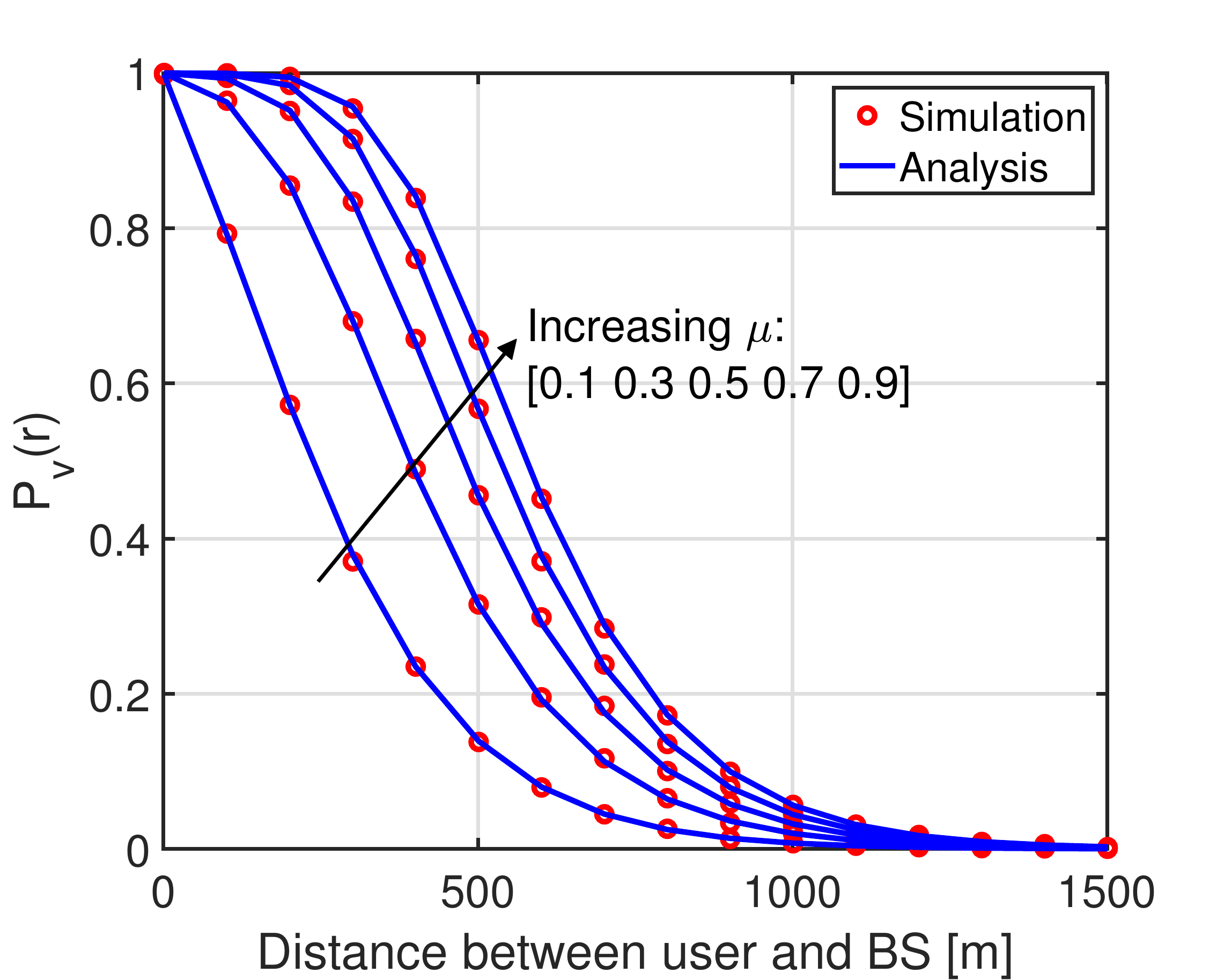}\label{fig1:c}}%
 \caption{The visibility probability $P_v(r)$, derived in Lemma~\ref{lemma3}, as a function of the distance $r$ between the user and the BS. The value of the density of blockages is $300$ km$^{-2}$ in (a), $500$ km$^{-2}$ in (b), and $700$ km$^{-2}$ in (c). }%
 \label{fig:plos}%
\end{figure}

In Fig.~\ref{fig:plos}, we plot the value of $P_v(r)$, derived in Lemma~\ref{lemma3}, for different values of the distance $r$ between the user and the BS. We observe that increasing the value of the fraction of RIS-equipped blockages $\mu$, significantly increases the visibility probability at lower values of the density of blockages (as seen in Fig~\ref{fig1:a}), while its influence reduces as the density of blockages increases (as seen in Fig~\ref{fig1:c}).

\begin{figure}
\centering
\includegraphics[width=0.5\columnwidth]{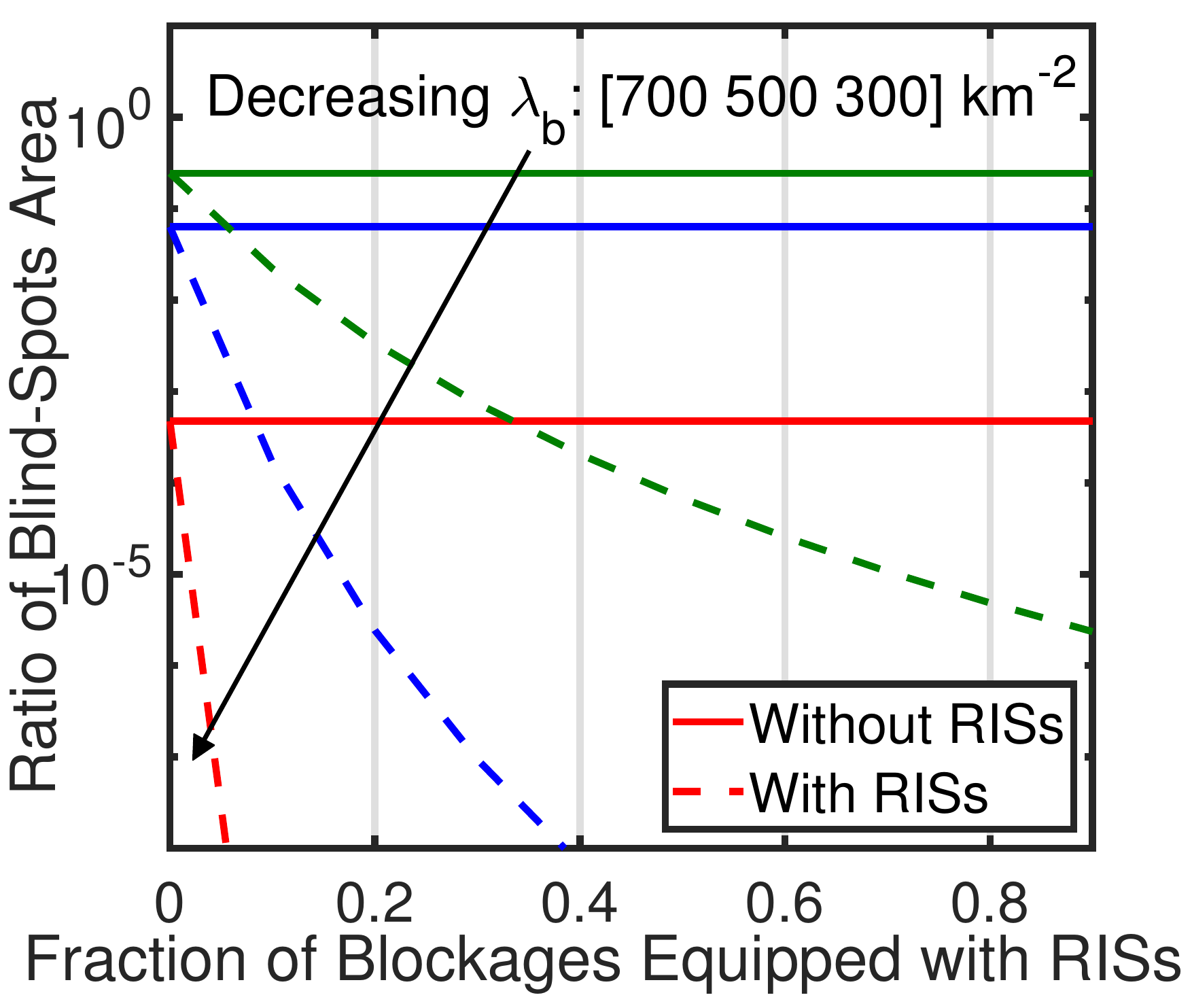}
\caption{ The average ratio of blind-spot areas reduces dramatically as we increase the ratio of RIS-equipped blockages $\mu$.}
\label{fig:not}
\end{figure}

In Fig.~\ref{fig:not}, we plot the value of $\mathcal{E}$ (the ratio of blind-spots to the total area), derived in Theorem~\ref{thm1}, for different values of $\mu$. We can observe that when the density of blockages are $300$ blockage/km$^2$, the value of $\mathcal{E}$ reduces to $10^{-5}$ by just equipping $2\%$ of the blockages with RISs (which means deploying $6$ RISs/km$^2$). However, as we increase the density of blockages to $700$ blockages/km$^2$, we observe that the required percentage of blockage to be equipped with RISs to reach the value of $\mathcal{E}=10^{-5}$ is $\mu=70\%$, which means deploying $490$ RISs/km$^2$. These values are useful to understand the required density of RISs to be deployed in different kinds of environments such as suburban areas with low density of blockages and high rise urban areas with high density of blockages.
\begin{figure}%
 \centering
 \subfloat[]{\includegraphics[width=0.45\columnwidth]{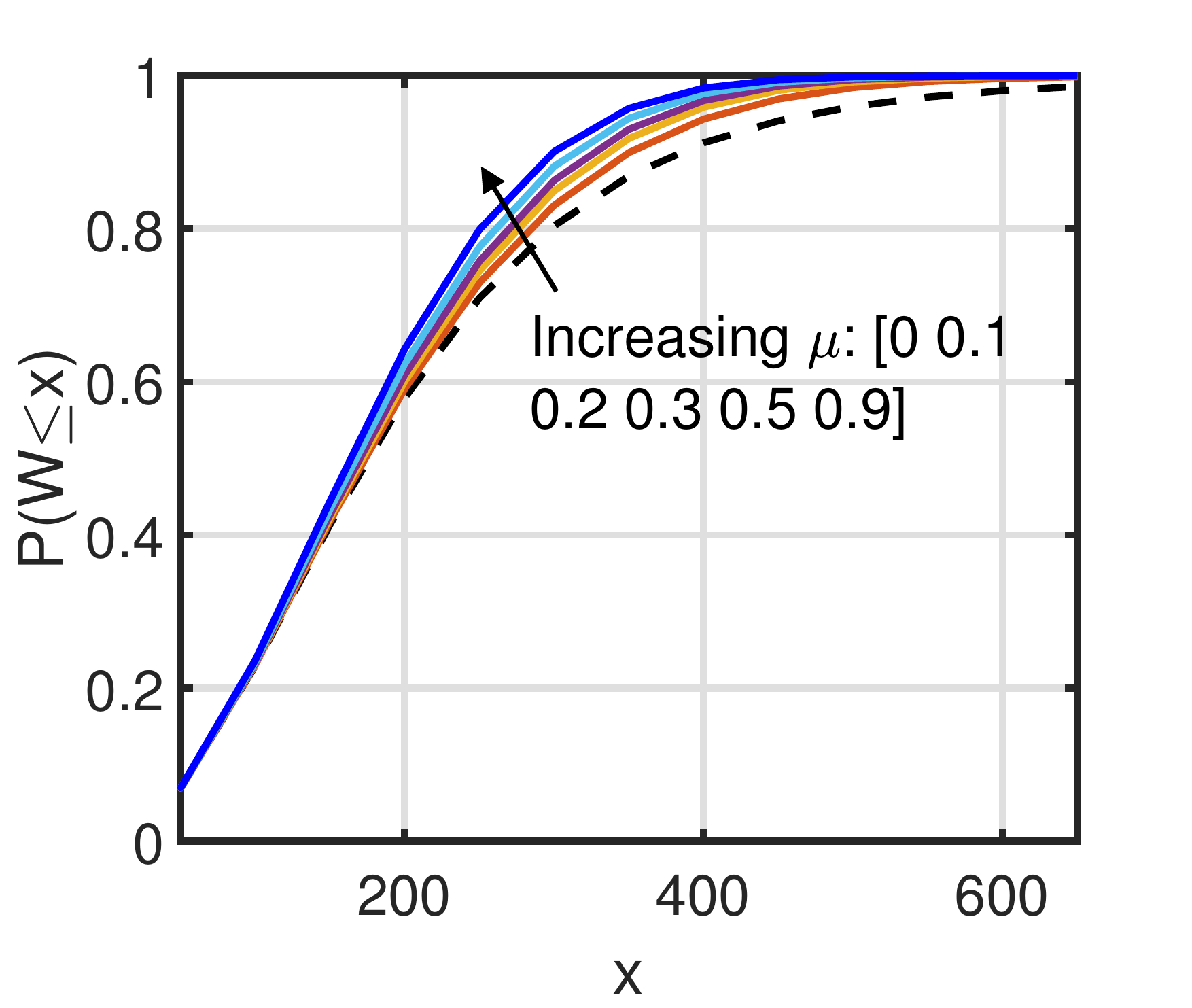}\label{fig:a}}%
 \subfloat[]{\includegraphics[width=0.45\columnwidth]{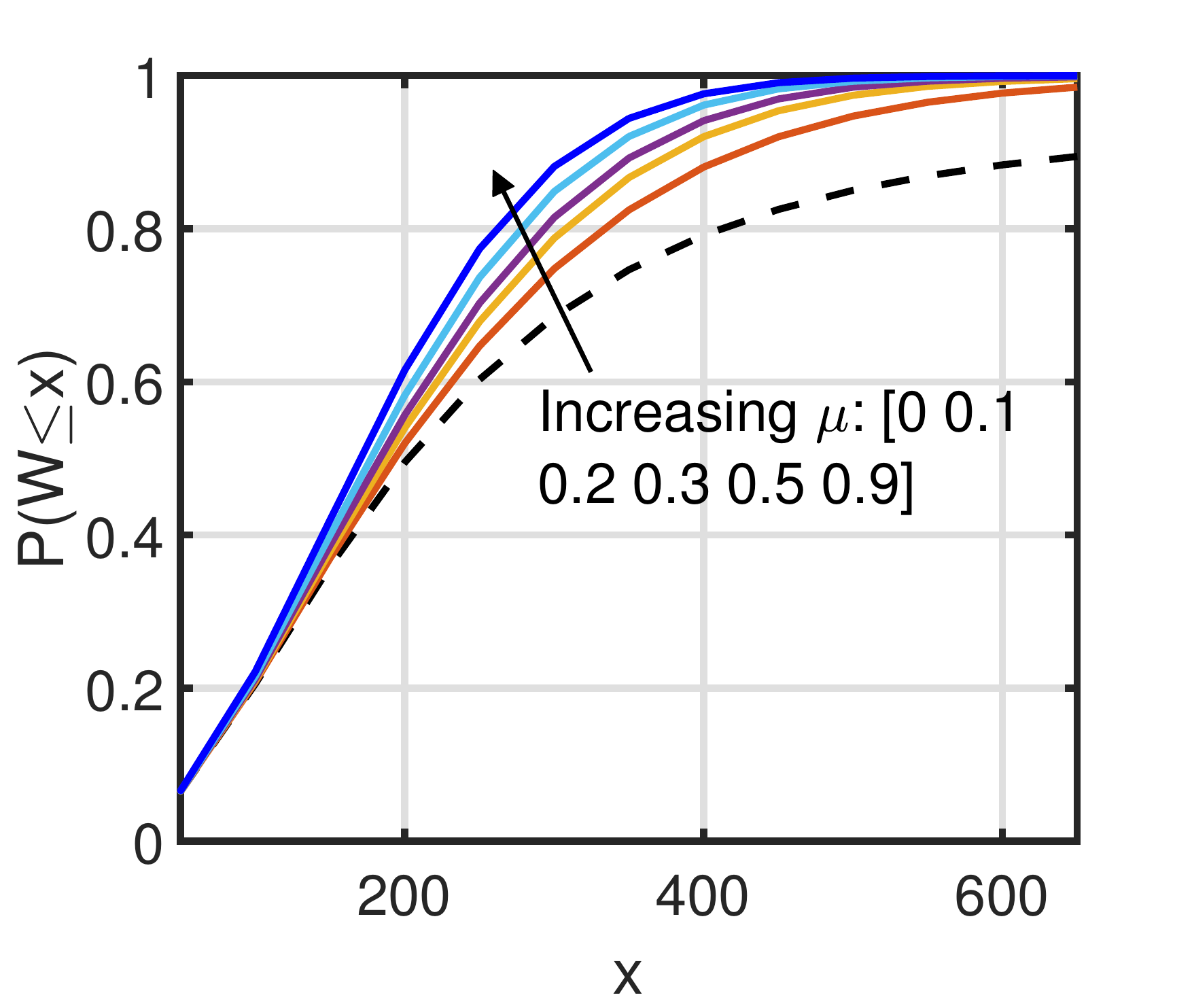}\label{fig:b}}\\
 \subfloat[]{\includegraphics[width=0.45\columnwidth]{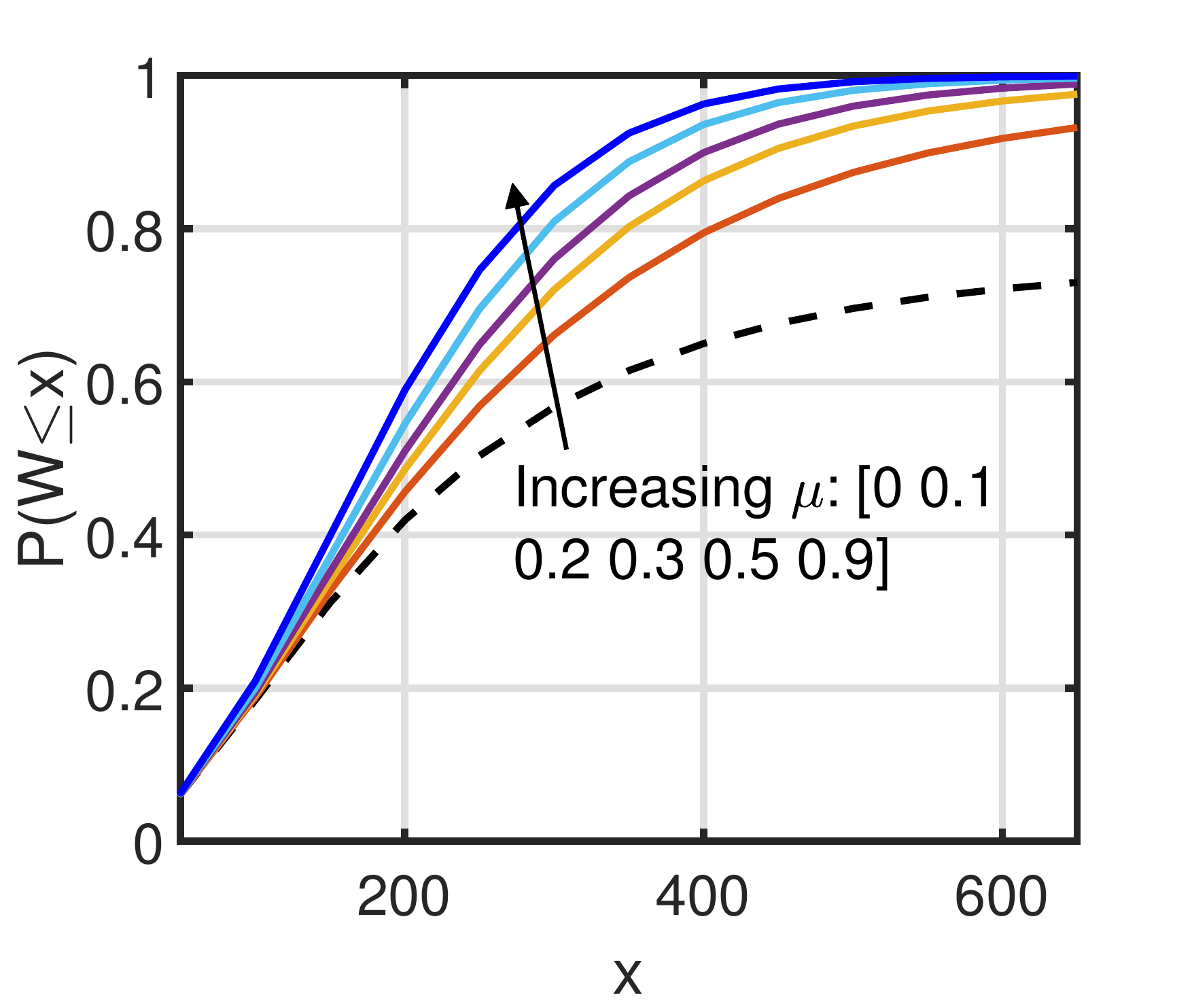}\label{fig:c}}%
 \caption{ The probability distribution of $W$, which is derived in Theorem~\ref{thm2}, for (a) $\lambda_b=300$ blockage/km$^2$, (b) $\lambda_b=500$ blockage/km$^2$, and (c) $\lambda_b=700$ blockage/km$^2$.}%
 \label{fig:cdf_W}%
\end{figure}

In Fig.~\ref{fig:cdf_W}, we plot the probability distribution of $W$ derived in Theorem~\ref{thm2} for different values of $\lambda_b$. Our comments in Remark~\ref{rem:W} can be observed by noticing that as $x$ tends to $\infty$, the value of $F_W(x)$ tends to $1-\mathcal{E}$, which was shown in Fig.~\ref{fig:not}.

\begin{figure}%
 \centering
 \subfloat[]{\includegraphics[width=0.45\columnwidth]{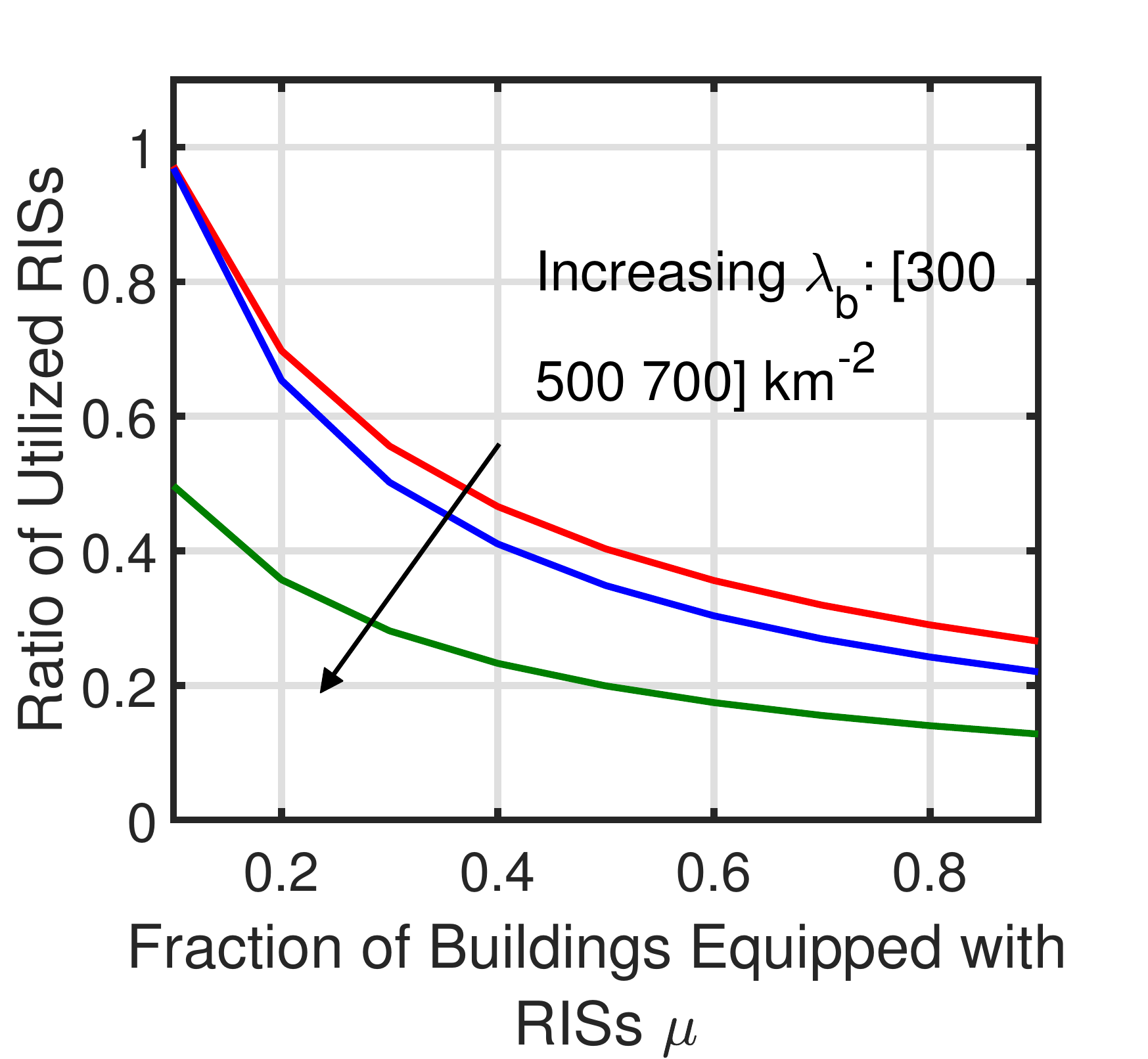}\label{fig:aaaa}}%
 \subfloat[]{\includegraphics[width=0.45\columnwidth]{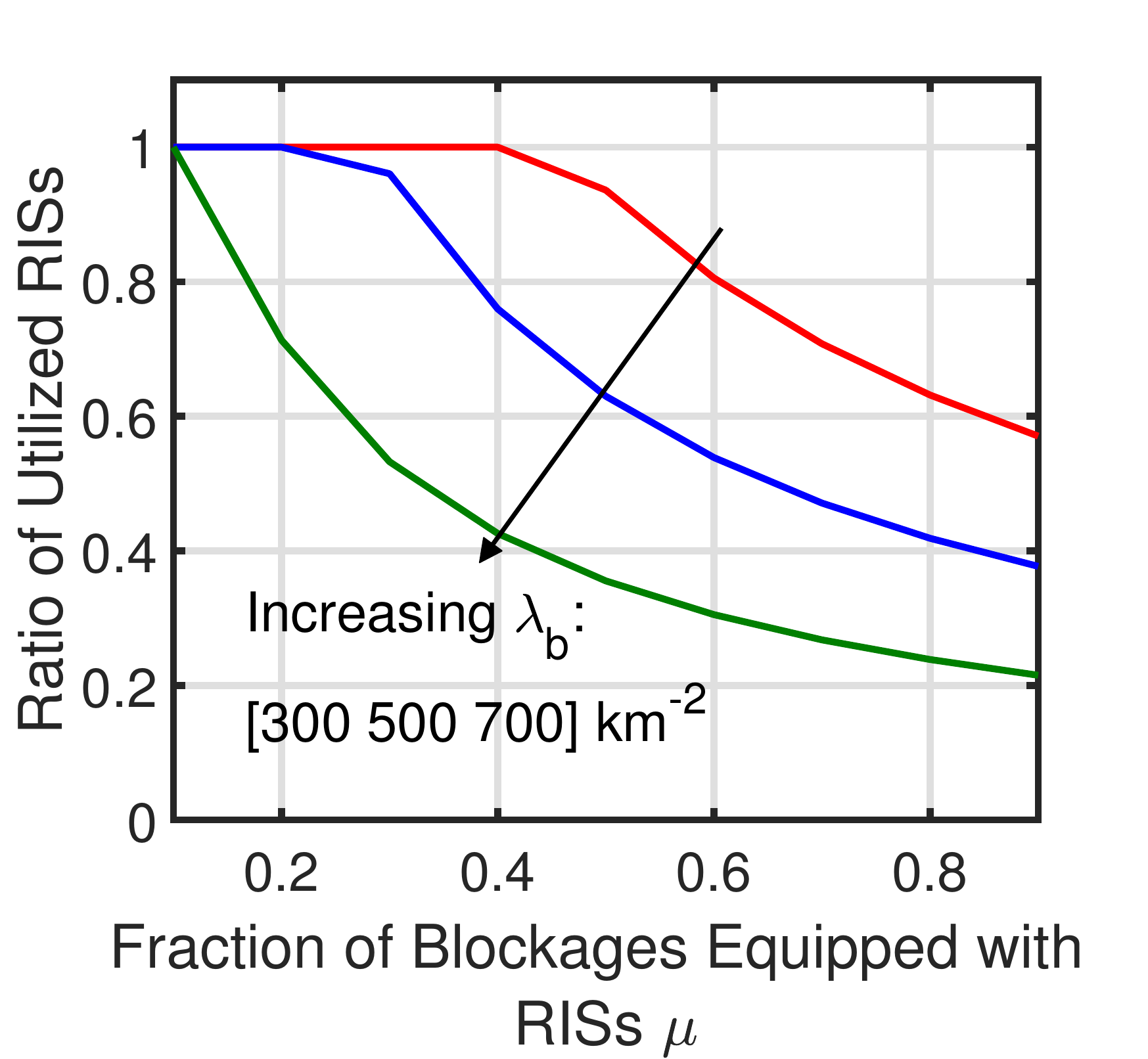}\label{fig:bbbb}}\\
  \subfloat[]{\includegraphics[width=0.45\columnwidth]{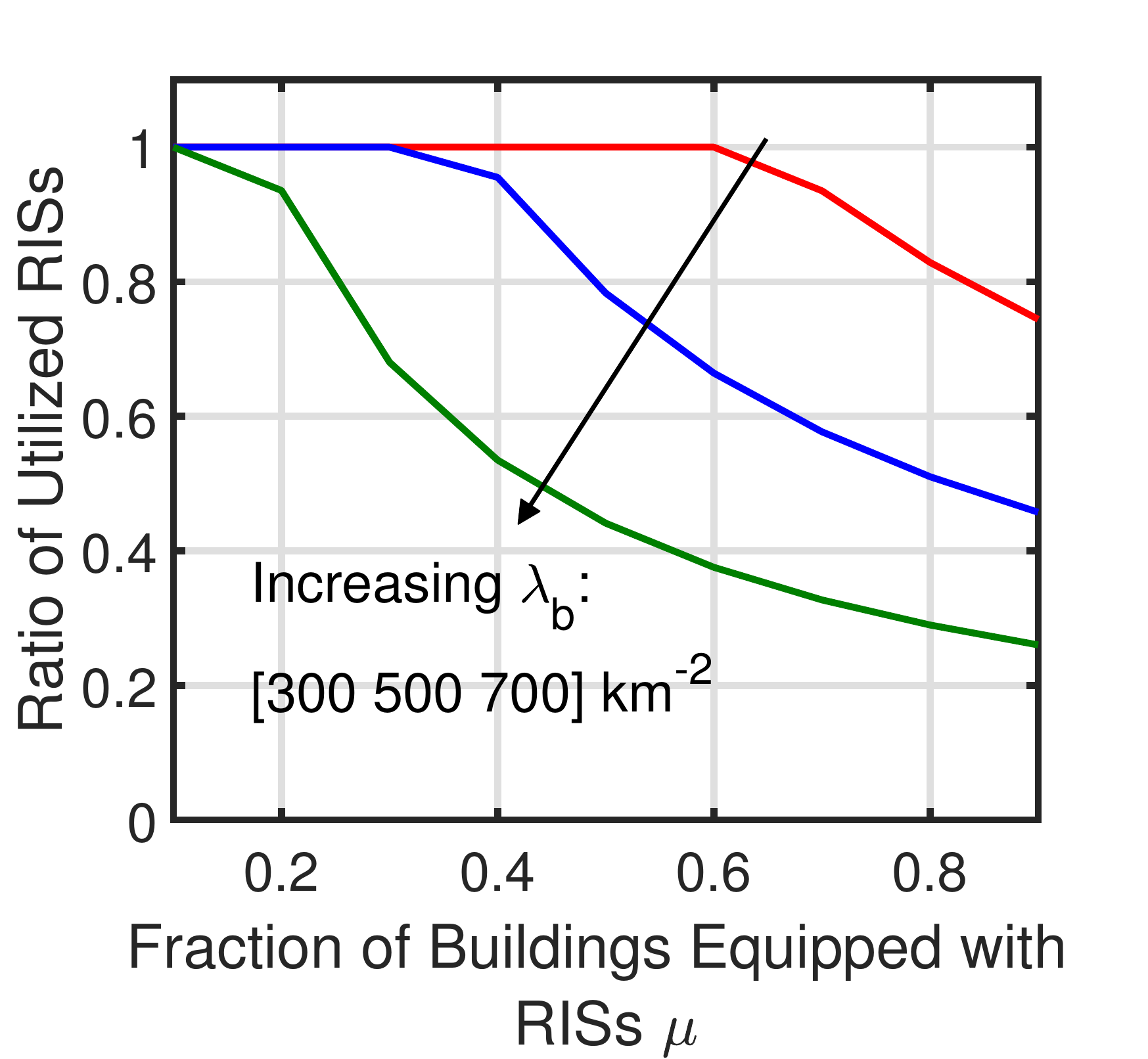}\label{fig:bbbb}}
 \caption{ The value of the deployment efficiency $\eta=\min\{1,\frac{\lambda_u\mathcal{A}_i}{\lambda_R}\}$, which was introduced in Remark~\ref{rem4}, for (a) $M=1$ meta-surface/RIS, (b) $M=2$ meta-surface/RIS, and (c) $M=3$ meta-surface/RIS.}%
 \label{fig:efficiency}%
\end{figure}

In Fig.~\ref{fig:efficiency}, we plot the efficiency $\eta=\min\{1,\frac{\lambda_u\mathcal{A}_i}{\lambda_R}\}$, for different values of $\lambda_b$. We notice that as we increase the value of $\mu$, the deployment efficiency goes down. This implies that as we increases the density of RISs, the actual fraction of utilized RISs decreases. This is mainly due to the unplanned, completely random deployment of RISs that we assume in this paper. These insights are of special importance since they provide an estimate for the required density of RISs when well-planned deployment is pursued to achieve similar performance to that of the random deployment. Recall that well-planned deployment means selecting the blockages at the most strategic locations and equipping them RISs. To sum up, the performance achieved by deploying $\mu\lambda_b$ RISs randomly without planned-deployment, can be achieved by well-planned deployment of only $\eta\mu\lambda_b$ RISs. Another observation from Fig.~\ref{fig:efficiency} is that the efficiency of deployment significantly increases as we increase the value of meta-surfaces per RIS.

\begin{figure}[!t]
    \centering
    \begin{minipage}{.45\textwidth}
        \centering
\includegraphics[width=1\columnwidth]{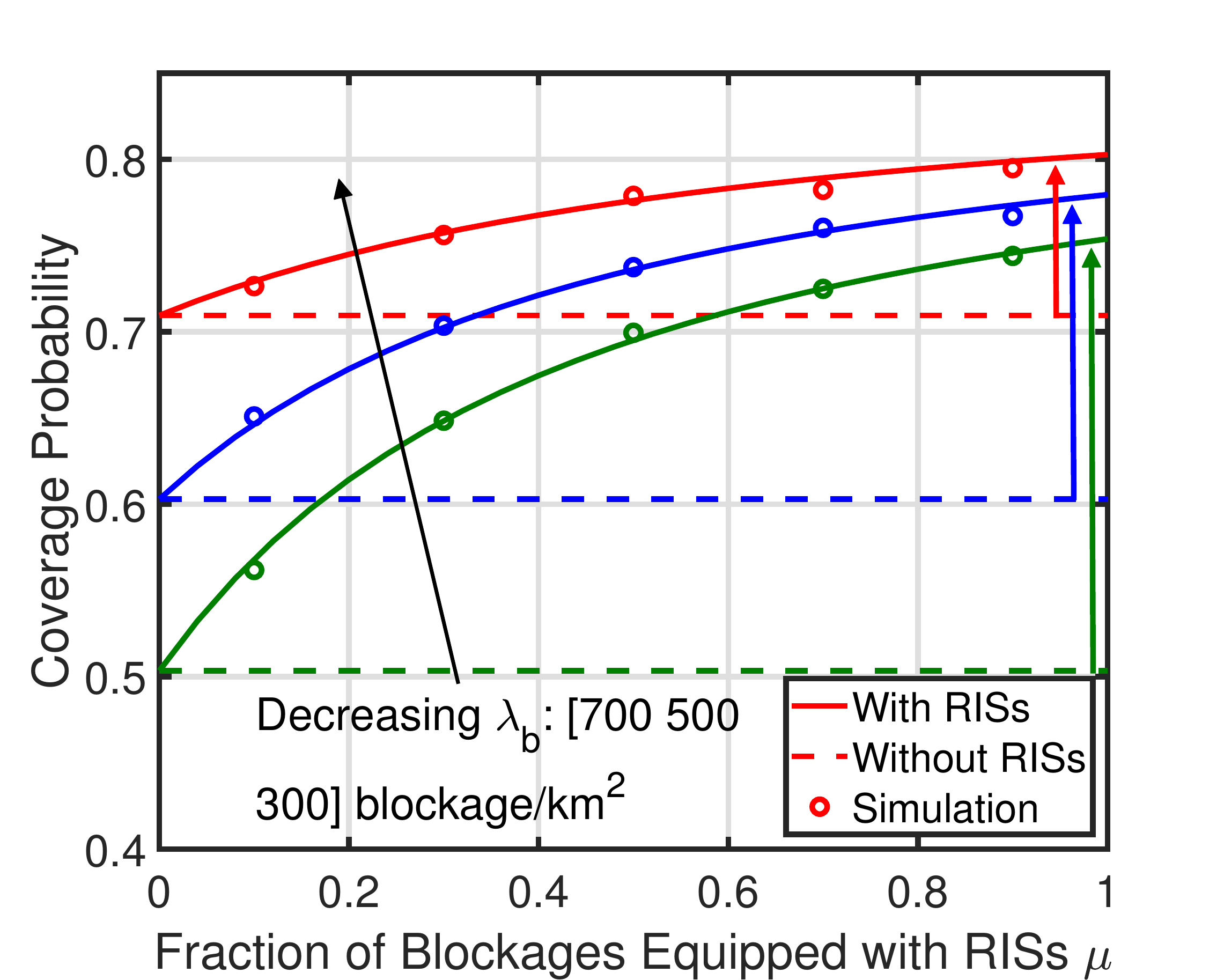}
\caption{The value of $P_{\rm cov}$, derived in Theorem~\ref{thm:cov}, against different values of $\mu$.}
\label{fig:cov_M_1}
    \end{minipage}%
    \hfill
    \begin{minipage}{0.45\textwidth}
        \centering
\includegraphics[width=1\columnwidth]{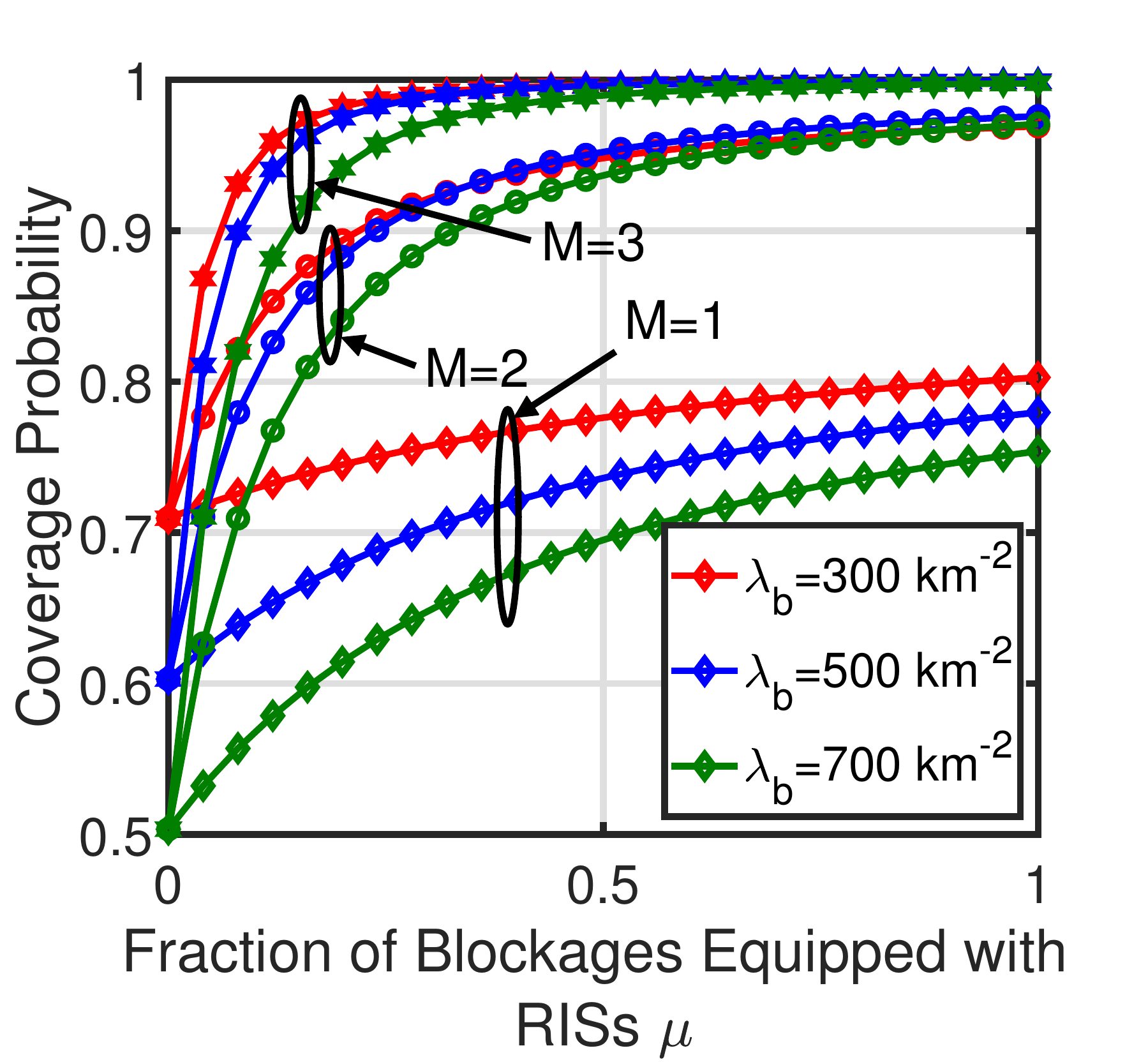}
\caption{The value of $P_{\rm cov}$ for different values of $\mu$ and $M$.}
\label{fig:cov_M_1_2_3}
    \end{minipage}
\end{figure}
In Fig.~\ref{fig:cov_M_1}, we plot the value of $P_{\rm cov}$, derived in Theorem~\ref{thm:cov}, for the case of $M=1$ against different values of $\lambda_b$ and $\mu$. The first insight drawn from this figure is the considerable enhancement in the value of $P_{\rm cov}$ when RISs are deployed. The second insight is that the deployment of RISs becomes more beneficial (larger increase in $P_{\rm cov}$) at high values of $\lambda_b$.

%

In Fig.~\ref{fig:cov_M_1_2_3}, we consider the case of having a fixed number of meta-surfaces $M$ per RIS. For that setup, we plot the value of $P_{\rm cov}$ for different values of $\mu$ and $M$. We observe the high influence of increasing the value of $M$ on the value of $P_{\rm cov}$. For the same system parameters, increasing the value of $M$ from $1$ to $3$ leads to an increase in the maximum achievable value of $P_{\rm cov}$ at $\mu=1$ from $0.8$ to $1$. On the same lines, we observe that, for instance at $\lambda_b=700$ blockage/km$^2$, the ratio of RIS-equipped blockages needed to achieve a value of $P_{\rm cov}=0.75$ reduces from $\mu=1$ when $M=1$ to $\mu=0.1$ when $M=2$ and to $\mu=0.05$ when $M=3$. These insights can be used to study the trade-off, in terms of cost, between deploying high density of RISs with low value of $M$ and deploying low density of RISs with high value of $M$.

\begin{figure}%
 \centering
 \subfloat[]{\includegraphics[width=0.45\columnwidth]{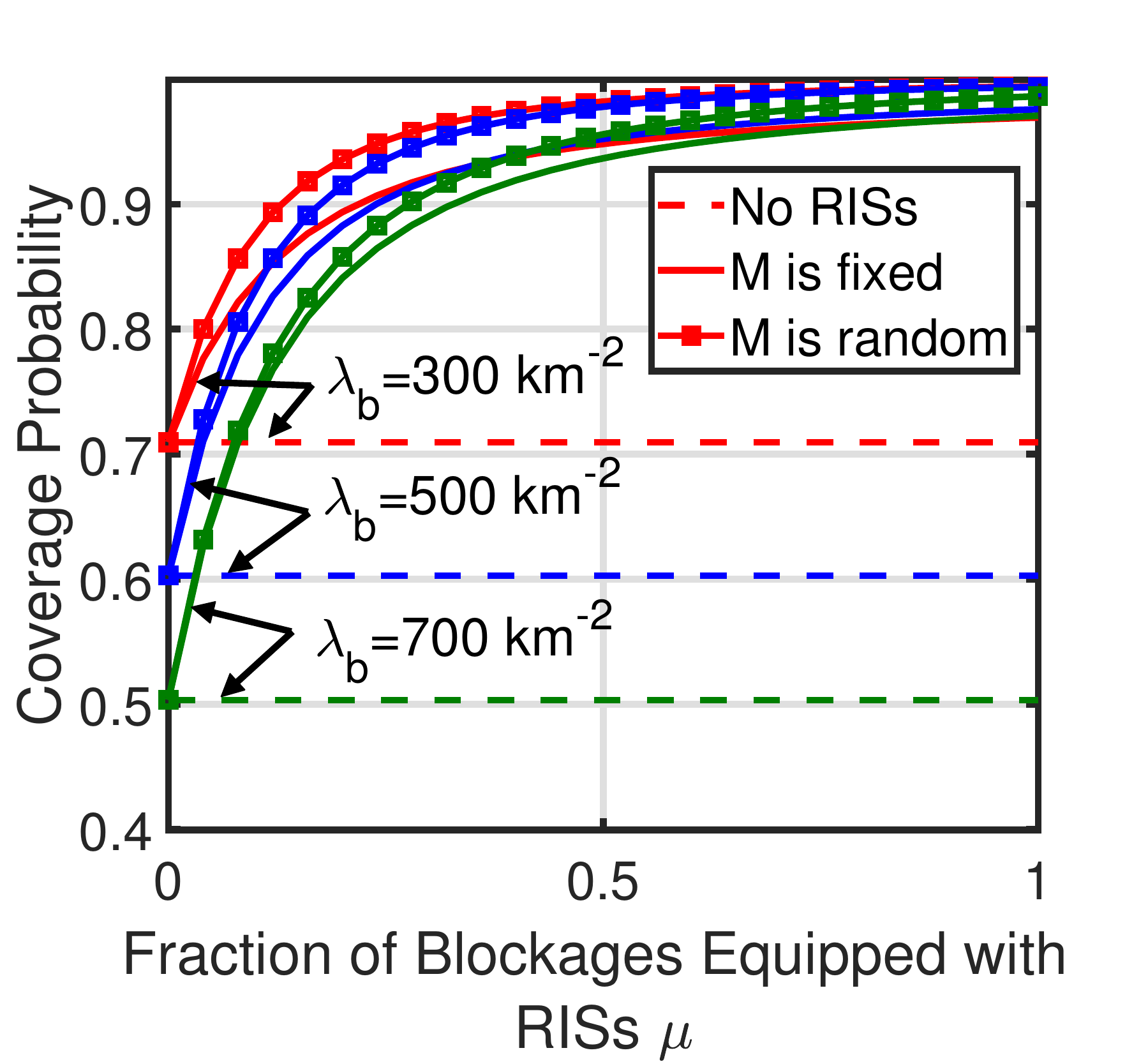}\label{fig:aaa}}
 \subfloat[]{\includegraphics[width=0.45\columnwidth]{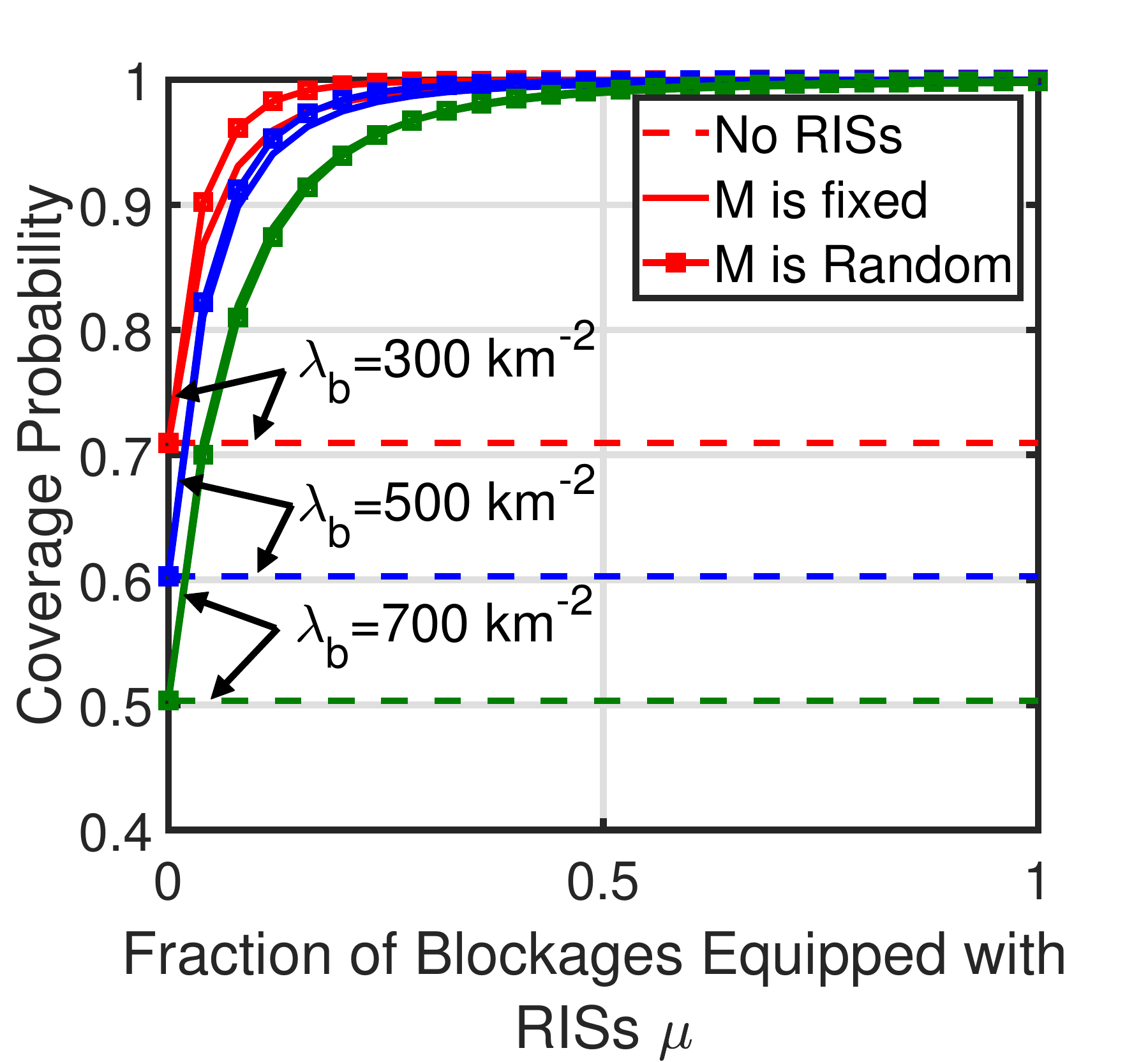}\label{fig:bbb}}
 \caption{ Comparing the value of $P_{\rm cov}$, derived in Theorem~\ref{thm:cov} for the cases of (i) fixed $M=M_F$ and (ii) uniformly distributed value of $M$ with $\mathbb{E}[M]=M_F$. In (a) we consider $M_F=2$ while in (b) we assume $M_F=3$.} 
 \label{fig:cov_fixed_random}%
\end{figure}

In Fig.~\ref{fig:cov_fixed_random}, we compare two scenarios: (i) the number of meta-surfaces per RIS is fixed $M=M_F$ and (ii) the number of meta-srufaces per RIS is uniformly distributed with $\mathbb{E}[M]=M_F$. We plot the values of $P_{\rm cov}$ against different values of $\mu$ for the two scenarios with $M_F=2$ in Fig.~\ref{fig:aaa} and $M_F=3$ in Fig.~\ref{fig:bbb}. We observe an increase in the value of $P_{\rm cov}$ when $M$ is uniformly distributed compared to the case of having a fixed $M$.
\section{Conclusions}
In this paper, we have provided the first stochastic geometry-based performance evaluation for large-scale deployment of RISs in cellular networks. We focused on a particular use case of the RISs, which is providing indirect LoS links for user-BS pairs that have blocked links. Modeling the locations of the BSs as a PPP and the blockages using line boolean model, we derived multiple important performance metrics. Firstly, for a given user-BS link, we derived the {\em indirect path probability}, defined by the probability of finding at least one RIS that can provide an indirect LoS link. Next, we derived the density of {\em visible BSs}, which is the density of BSs that have an LoS (direct or indirect) with the typical user. We then used this density to derive the area of {\em blind-spots}. Considering a random deployment scenario, where there is no specific criteria for selecting the blockages that will be coated with RISs, we derived the deployment efficiency of RISs. We used this efficiency to emphasize on the importance of well-planned deployment in terms of reducing the required deployment density to achieve the required performance levels. Finally, we derived the probability that the average path-loss is below a predefined threshold. We showed that the number of meta-surfaces per RIS has a significant effect on the performance of the cellular network. 

The drawn system-level insights from the analysis in this paper can be beneficial for the pre-deployment design and planning of RIS-enabled systems. For instance, we provided a useful guideline on the required density of RISs for different values of blockage density. In addition, we provided insights on the performance limitations of well-planned deployment of RISs, which is the deployment process that smartly selects the locations that has the highest effect on the system performance.

This work can be extended in many directions. Firstly, the provided framework can be used to study the influence of RIS deployment on multiple aspects of the performance of the cellular network. For instance, it can be used to study RIS-enabled localization, secrecy-enhancement, and wireless power transfer in large-scale networks. Furthermore, while we provided useful comments on the limitations of well-planned deployment of RISs, the deployment criteria that could achieve these limitations are still an open research problem. In particular, for the considered system, and for a given value of the fraction RIS-equipped blockages, it is still needed to compute the optimal subset of blockages to be equipped with RISs.
\appendices
\section{Proof of Lemma~\ref{lemma1}}\label{app:lemma1}
For the setup provided in Fig.~\ref{fig:proof1}, the angle $\theta$ between the RIS and the user-RIS link needs to be greater than or equal $\psi$ in order to be able to provide an indirect path between the user and the BS. First, we compute the value of $d_{I-B}$ in terms of $t$, $r$, and $\phi$, which can be achieved using the cosine law as follows
\begin{align}\label{a1}
d_{I-B}^2=r^2+t^2-2rt\cos(\phi).
\end{align}
Similarly, the value of $\psi$ can be derived in terms of $r$, $t$, and $\phi$ as follows
\begin{align}
r^2&=t^2+d_{I-B}^2-2td_{I-B}\cos(\psi)\nonumber\\
&\overset{(a)}{=}t^2+r^2+t^2-2rt\cos(\phi)-2t\sqrt{r^2+t^2-2rt\cos(\phi)}\cos(\psi),\\
\therefore \cos(\psi)&=\frac{t-r\cos(\phi)}{\sqrt{r^2+t^2-2rt\cos(\phi)}},\label{a2}
\end{align}
where $(a)$ follows by replacing for $d_{I-B}$ using (\ref{a1}). In order to ensure the capability of the RIS to provide an indirect path between the user and the BS, three conditions need to be satisfied:
\begin{itemize}
\item The user must be facing the RIS-equipped side of the blockage.
\item Both the user and the BS need to be on the same side of the RIS.
\item The RIS has an LoS-link with the user.
\item The RIS has an LoS-link with the BS.
\end{itemize} 
The first condition is satisfied with probability $\frac{1}{2}$. The second condition is satisfied only when $\theta\geq\psi$. Hence, the probability that the RIS satisfies the first condition is
\begin{align}
\mathcal{C}(r,t,\phi)&=\mathbb{P}(\theta\geq\psi)\nonumber\\
&\overset{(b)}{=}\mathbb{P}\left(\theta\geq\cos^{-1}\left(\frac{t-r\cos(\phi)}{\sqrt{r^2+t^2-2rt\cos(\phi)}}\right)\right)\nonumber\\
&\overset{(c)}{=}1-\frac{1}{\pi}\cos^{-1}\left(\frac{t-r\cos(\phi)}{\sqrt{r^2+t^2-2rt\cos(\phi)}}\right),
\end{align}
where $(b)$ follows by replacing for $\psi$ using (\ref{a2}), and $(C)$ is due to $\theta\sim U(0,\pi)$. Given that the third condition is satisfied with probability $P_{\rm LoS}(t)$ and the fourth condition is satisfied with probability $P_{\rm LoS}(d_{I-B})$, the final expression in Lemma~\ref{lemma1} follows.
\section{Proof of Lemma~\ref{lemma2}}\label{app:lemma2}
Applying Proposition~\ref{prop2}, we assume that the typical user is located at the origin and the BS is located at a distance $r$ from the origin on the positive direction of the $x$-axis. Hence, the polar coordinates of each point in $\Psi_R$ are $t$ and $\phi$. Based on the results in Lemma~\ref{lemma1}, the locations of the RISs that are capable of providing an indirect path between the typical user and a BS at distance $r$ are modeled by an inhomogeneous PPP $\Psi_R^{I|r}$ with density $\lambda_I(r,t,\phi)=\lambda_Ra_i(r,t,\phi)$. Hence, the probability of having at least one indirect path between the typical user and a BS at distance $r$ is
\begin{align}\label{a21}
P_I(r)=\mathbb{P}(\mathcal{N}_{\Psi_R^{I|r}}(\mathbb{R}^2)>0)=1-\mathbb{P}(\mathcal{N}_{\Psi_R^{I|r}}(\mathbb{R}^2)=0),
\end{align}
where $\mathcal{N}_{\Psi_R^{I|r}}(A)$ is the number of points of $\Psi_R^{I|r}$ that lie inside the area $A$. Given that $\Psi_R^{I|r}$ is an inhomogeneous PPP with density $\lambda_I(r,t,\phi)$, then
\begin{align}\label{a22}
\mathbb{P}(\mathcal{N}_{\Psi_R^{I|r}}(\mathbb{R}^2)=0)=\exp\left(-\int_{\mathbb{R}^2}\lambda_I(r,t,\phi)t{\rm d}t{\rm d}\phi\right).
\end{align}
Substituting (\ref{a22}) in (\ref{a21}) leads to the final expression in Lemma~\ref{lemma2}.
\section{Proof of Lemma~\ref{lemma4}}\label{app:lemma4}
For a given LIS, and a BS at distance $r$ from the typical user, it can be observed from Fig.~\ref{fig:proof1} that the length of the indirect path through the RIS is $t+d_{I-B}$. Using (\ref{a1}), this length is equal to $t+\sqrt{t^2+r^2-2rt\cos(\phi)}$. Hence, for a given $r$, the probability that the shortest indirect path $R_{i,k}$ is less than $x$ is
\begin{align}
F_{R_{i,k}|r}(x)&=\mathbb{P}(R_{i,k}\leq x|r)\nonumber\\
&=1-\mathbb{P}(R_{i,k}\geq x|r)\nonumber\\
&=1-\mathbb{P}\left(\mathcal{N}_{\Psi_{R,k}^{I|r}}(\mathcal{S})=0\right)\nonumber\\
&\overset{(c)}{=}1-\exp\left(-\int_\mathcal{S}\lambda_{I,k}(r,t,\phi)) t{\rm d}t{\rm d}\phi\right),
\end{align}
where $\Psi_{R,k}^{I|r}$ models the locations of the RISs that have $k$ meta-surfaces and are capable of providing an indirect LoS-link between a user and a BS at distance $r$, $\lambda_{I,k}(r,t,\phi)=\rho_k \lambda_{I}(r,t,\phi)$ is the density of $\Psi_{R,k}^{I|r}$, $\mathcal{S}=\{t,\phi:t+\sqrt{t^2+r^2-2rt\cos(\phi)}<x\}$, and $(c)$ follows from the fact that, as explained in Appendix~\ref{app:lemma2}, $\Psi_{R,k}^{I|r}$ is an inhomogeneous PPP with density $\lambda_{I,k}(r,t,\phi)=\rho_k\lambda_R a_i(r,t,\phi)$. The final expression follows after rewriting $\mathcal{S}$ as
\begin{align}
\mathcal{S}=\left\{t,\phi:t<\frac{x^2-r^2}{2(x-r\cos(\phi))}\right\}.
\end{align}
The distribution of the shortest path $R_i$ using any RIS can be derived as follows
\begin{align}
\mathbb{P}(R_i\leq x|r)&=1-\mathbb{P}(R_i\geq x|r)\nonumber\\
&=1-\prod_{k\in\mathcal{M}}\mathbb{P}(R_{i,k}\geq x|r),
\end{align}
which leads to the final expression in the lemma.
\section{Proof of Lemma~\ref{lemma:Ri}}\label{app:lemma:Ri}
Based on the preliminary result in Proposition~\ref{prelim:lem1}, we can use the results in (\ref{cdf_cond1}) and (\ref{cdf_cond2}) as thinning probabilities for $\Psi_{BS}$, leading to inhomogeneous PPPs $\Psi_{BS}^{I,k|x}$ and $\Psi_{BS}^{I|x}$, with densities $\lambda_{BS}P_{\rm NLoS}(r)F_{R_{i,k}|r}(x)$ and $\lambda_{BS}P_{\rm NLoS}(r)F_{R_i|r}(x)$, respectively. Hence, the probability distribution of $R_{i,k}$ is
\begin{align}
F_{R_{i,k}}(x)=1-\mathbb{P}(\mathcal{N}_{\Psi_{BS}^{I,k|x}}(\mathbb{R}^2)=0),
\end{align}
where $\mathcal{N}_{\Psi_{BS}^{I,k|x}}(\mathbb{R}^2)$ is the number of points of $\Psi_{BS}^{I,k|x}$ in $\mathbb{R}^2$. Similar approach can be used to derive $F_{R_i}(x)$.
\section{Proof of Theorem~\ref{thm2}}\label{app:thm2}
The shortest path between the typical user and a visible BS can either be direct or indirect. Hence, the CDF of the shortest path length $W$ can be derived as follows
\begin{align}
\mathbb{P}(W\leq x)&=1-\mathbb{P}(W> x)\nonumber\\
&=1-\mathbb{P}(\mathcal{N}_{\Psi_{BS}^{d}}(0,x)=0,\mathcal{N}_{\Psi_{BS}^{I|x}}(\mathbb{R}^2)=0),
\end{align}
where $\Psi_{BS}^{d}$ is the inhomogeneous PPP modeling the locations of the BSs that have LoS with the typical user, with density $\lambda_{BS}P_{\rm LoS}(r)$. The inhomogeneous PPP $\Psi_{BS}^{I|x}$ presents the locations of the BSs that have an NLoS with the typical user but at least one indirect path with length less than $x$. The density of $\Psi_{BS}^{I|x}$ is $\lambda_{BS}P_{\rm NLoS}(r)F_{R_i|r}(x)$. Given that the events $\mathcal{N}_{\Psi_{BS}^{d}}(0,x)=0$ and $\mathcal{N}_{\Psi_{BS}^{I|x}}(\mathbb{R}^2)=0$ are independent, we get 
\begin{align}
\mathbb{P}(W\leq x)&=1-\mathbb{P}(\mathcal{N}_{\Psi_{BS}^{d}}(0,x)=0)\mathbb{P}(\mathcal{N}_{\Psi_{BS}^{I|x}}(\mathbb{R}^2)=0),
\end{align}
which leads to the final expression in Theorem~\ref{thm2}.
\section{Proof of Theorem~\ref{thm3}}\label{app:thm3}
Recalling Sec.~\ref{sec:asso}, conditioned on $R_d$, a BS at distance $r$ from the typical user provides a path-loss lower than that provided by the nearest direct LoS BS with probability \begin{align} P_{\rm NLoS}(r)\left(1-\prod_{k\in\mathcal{M}}\bar{F}_{R_{i,k|r}}\left(k^\frac{2}{\alpha}R_d\right)\right).
\end{align}
Hence, applying Proposition~\ref{prelim:lem1}, the probability that the typical user associates with a BS through a direct LoS is
\begin{align}
\mathcal{A}_d=\mathbb{E}_{R_d}\left[\exp\left(-2\pi\lambda_{BS}\int_0^\infty P_{\rm NLoS}(r)\left(1-\prod_{k\in\mathcal{M}}\bar{F}_{R_{i,k|r}}\left(k^\frac{2}{\alpha}R_d\right)\right)r{\rm d}r\right)\right].
\end{align}
Taking the expectation over $R_d$ and recalling that $\mathcal{A}_i=1-\mathcal{A}_d-\mathcal{E}$ leads to the final expression.
\section{Proof of Theorem~\ref{thm:cov}}\label{app:thm:cov}
For a given BS at a distance $r$ from the origin, the probability that this BS is capable to provide a path-loss less than $\tau$ to the typical user is
\begin{align}
P_{{\rm cov}|r} =\left\{
	\begin{array}{ll}
		P_{\rm LoS}(r)+P_{\rm NLoS}(r)\left(1-\prod_{k\in\mathcal{M}}\bar{F}_{R_{i,k}|r}\left((\tau k^2)^\frac{1}{\alpha}\right) \right) & \mbox{if } {r}\leq \tau^\frac{1}{\alpha} \\
		P_{\rm NLoS}(r)\left(1-\prod_{k\in\mathcal{M}}\bar{F}_{R_{i,k}|r}\left((\tau k^2)^\frac{1}{\alpha}\right) \right) & \mbox{if } {r}\geq \tau^\frac{1}{\alpha} 
	\end{array}.
	\right.
\end{align}
Applying Proposition~\ref{prelim:lem1} by using $P_{{\rm cov}|r}$ as a thinning probability for $\Psi_{BS}$, we get
\begin{align}
P_{\rm cov}=1-\exp\left(-2\pi\lambda_{\rm BS}\int_0^\infty P_{{\rm cov}|r} r{\rm d} r\right).
\end{align}
After simple algebraic manipulations, the final result in Theorem~\ref{thm:cov} can be achieved.

\bibliographystyle{IEEEtran}
\bibliography{Draft_v0.19.bbl}
\end{document}